\documentclass[%linenumbers,
modern,astrosymb]{aastex7}

\usepackage{amsmath}

\usepackage{hyperref}

\def\subparagraph{\@startsection{subparagraph}{5}{\z@}%
  {1ex plus 1ex minus .2ex}{-0.5\parindent}{\small\it}}

\PassOptionsToPackage{draft}{graphicx}
\begin{document}

\title{Biosignature False Positives in Potentially Habitable Planets around M-dwarfs: The Effect of UV Radiation from One Flare.}

\author[orcid=0000-0002-2857-0131,gname=Arturo,sname='Miranda-Rosete']{Arturo Miranda-Rosete}
\affiliation{Instituto de Ciencias Nucleares, Universidad Nacional Aut\'onoma de M\'exico}
\email[show]{arturo.miranda@correo.nucleares.unam.mx}  

\author[orcid=0000-0002-2240-2452,gname=Ant\'igona, sname='Segura']{Ant\'igona Segura}
\affiliation{Instituto de Ciencias Nucleares, Universidad Nacional Aut\'onoma de M\'exico}
\affiliation{NASA Nexus for Exoplanet System Science, Virtual Planetary Laboratory Team, Seattle, WA}
\email{antigona@nucleares.unam.mx}

\author[0000-0002-2949-2163]{Edward W. Schwieterman}
\affiliation{NASA Nexus for Exoplanet System Science, Virtual Planetary Laboratory Team, Seattle, WA}
\affiliation{Department of Earth and Planetary Sciences, University of California, Riverside, CA}
\affiliation{Blue Marble Space Institute of Science, Seattle, WA, USA}
\email{eschwiet@ucr.edu}


\begin{abstract}

Many past studies have predicted the steady-state production and maintenance of abiotic O\textsubscript{2} and O\textsubscript{3} in the atmospheres of CO\textsubscript{2}-rich terrestrial planets orbiting M-dwarf stars. However, the time-dependent responses of these planetary atmospheres to flare events---and the possible temporary production or enhancement of false-positive biosignatures therein---has been comparatively less well studied. Most past works that have modeled the photochemical response to flares have assumed abundant free oxygen like that of the modern or Proterozoic Earth. Here we examine in detail the photochemical impact of the UV emitted by a single flare on abiotic O\textsubscript{2}/O\textsubscript{3} production in prebiotic, CO\textsubscript{2}-dominated atmospheres of M-dwarf planets with CO\textsubscript{2} levels ranging from 3\% to 80\% of 1 bar. We find that a single flare generally destroys O\textsubscript{2} while modestly enhancing their column densities over intermediate timescales. We simulate the spectral observables of both the steady-state atmosphere and time-dependent spectral response over the flare window for both emitted and transmitted light spectra. Over the course of the flare, the O\textsubscript{3} UV Hartley band is decreased by a maximum of 47 ppm. In both emitted and transmitted light spectra, the 9.65~$\mu$m O\textsubscript{3} band is hidden by the overlapping 9.4~$\mu$m CO\textsubscript{2} band for all scenarios considered. Overall, we find that the possible enhancements of abiotic O\textsubscript{3} due to a single flare are small compared to O\textsubscript{3}’s sensitivity to other parameters such as CO\textsubscript{2} and H\textsubscript{2}O abundances or the availability of reducing gases such as H\textsubscript{2}. 

\end{abstract}



%% cross-references to sections, equations, tables, and fi gures.

\section{Introduction} \label{sec:intro}

M-dwarf star systems are compelling targets in the search for habitable planets due their preponderance in the solar neighborhood \citep{bochanski_luminosity_2010}, long main-sequence lifetime on the order of ~10\textsuperscript{11} yr \citep{adams_dying_1997}, and the large numbers of Earth-sized planets already detected in these systems \citep{mulders_increase_2015,garrett_planet_2018}.

For exoplanets, planetary habitability is often defined as the potential to sustain surface life, and thus requires the presence of surface water and an atmosphere \citep[e.g.][]{segura_search_2010}. Surface life, and an interface between liquid water and the atmosphere, would allow for the exchange of gases and facilitate the potential detectability of remote biosignatures \citep{kasting_remote_2014}. Therefore, we often limit the search of life to planets around stars in the main sequence that can sustain surface liquid water even though other habitable environments likely exist elsewhere. Such planets should be rocky, that is, primarily composed of iron and silicates, lie inside the circumstellar habitable zone, and possess an atmosphere with sufficient greenhouse gases to maintain a clement environment \citep[e.g.][]{domagal-goldman_astrobiology_2016}. 

Planets orbiting M-dwarfs present an opportunity in the search of life due to the observational advantages that their host stars present: first is the sheer number of M-dwarfs (76.45\% of main-sequence stars in the solar neighborhood; \citeauthor{bochanski_luminosity_2010}, \citeyear{bochanski_luminosity_2010}), second the ratio between planetary and stellar masses means a larger Doppler shift in radial velocity observations, and third the ratio between planetary and stellar radii causes a deeper depths in the light curves measured during planetary transits \citep{gould_sensitivity_2003, nutzman_design_2008}. Furthermore, the low stellar luminosity causes the habitable zone to be much closer to M-dwarf stars than for the Sun or similar stars, which in turn increases the geometrical probability of transit and thus of detection. With the caveat that stellar activity will be an unavoidable challenge for planet detection with both methods \citep[e.g.][]{newton_impact_2016,clarice2024}.

Our current strategy for life detection defines a biosignature as any element, molecule, substance or characteristic that could be used as evidence of life whether present or past and different from an abiotic background \citep{des_marais_nasa_2008, hays_nasa_2015, schwieterman_exoplanet_2018}. In the search for and detection of atmospheric biosignatures it is of paramount importance to understand the possible abiotic mechanisms which may produce the same products of life, particularly because the study of life outside the solar system is forced to use indirect methods \citep{catling_exoplanet_2018}. Here we are interested in atmospheric O\textsubscript{2} and O\textsubscript{3}, which on Earth are considered unequivocal evidence for life, as the Earth system lacks an abiotic mechanism to produce the modern amount of O\textsubscript{2} or O\textsubscript{3} \citep{meadows_reflections_2017, meadows_exoplanet_2018}. However, this may not be the case of other planets, since possible false-positive biosignatures of O\textsubscript{2} and O\textsubscript{3} have been found to be produced by three mechanisms: (i) low non-condensable gas inventories \citep{wordsworth_abiotic_2014}, (ii) enhanced M-dwarf pre-main sequence stellar luminosity \citep{tian_high_2014, luger_habitable_2015}, and (iii)) stellar-spectrum-driven photochemical production \citep{selsis_signature_2002, grenfell_sensitibity_2014, tian_high_2014, gao_stability_2015, wordsworth_redox_2018}. The latter mechanism is relevant for planets in the habitable zone of M-dwarfs because they receive more far-UV (FUV) radiation in the wavelength range where CO\textsubscript{2} is photolyzed ($<$ 200 nm), producing oxygen, which can lead to the generation of O\textsubscript{2} and O\textsubscript{3}. At the same time, M-dwarfs produce less NUV radiation ($>$ 200 nm) that is crucial for driving the catalytic cycles that lead to the recombination of CO and O back into CO\textsubscript{2} \citep{gao_stability_2015, harman_abiotic_2015, ranjan_photochemistry_2020}.

As generally active stars, M-dwarfs generate continuous emissions of high-energy radiation and particles from their chromosphere, such as flares. Flares involve an increment of the energy emitted by the star in all wavelengths, from X-rays to radio. These bursts of energy are divided in two phases: the initial phase called impulsive, which lasts less than 30 minutes until the flare reaches its maximum emission; and the gradual phase, that can last a few hours for the most energetic flares. The increase in UV emission (100 nm $<$ $\lambda$ $<$ 350 nm) during flares has been shown to have an impact on the photochemistry of exoplanets’ atmospheres in the habitable zone, including the abundance of possible biosignatures \citep{segura_effect_2010, tilley_modeling_2019, chen_persistence_2021}. 

Works related to the impact of the UV emission by flares on planets in the habitable zone of M-dwarfs have focused either on the survival of life exposed to this radiation on the surface \citep{rugheimer_effect_2015, omalley-james_uv_2017, vida_frequent_2017, estrela_superflare_2018, estrela_surface_2020}, or the potential of the UV radiation  to drive the chemistry necessary for the origins of life \citep{buccino_uv_2007, Ranjan_2017, rimmer_origin_2018, armas-vazquez_impact_2023}.

Our work builds upon previous efforts by \citet{segura_effect_2010} and \citet{tilley_modeling_2019}. \citet{segura_effect_2010} studied the effect of one large flare emitted by a dM3e very active star, AD Leonis (AD Leo), on the ozone chemistry of a planetary atmosphere similar to present Earth’s atmosphere. They found that ozone is barely depleted ($<$ 1\%) during the flare, which means life would be protected from UV emitted during this flare. \citet{tilley_modeling_2019} extended the work of \citet{segura_effect_2010} simulating the effects of long periods of flare activity in a present Earth-like planet’s atmospheric chemistry while unprotected by an intrinsic magnetic field, showing that the ozone column can be depleted up to a 94\% in a 10 yr period by stellar activity if the effect of high-energy particles emitted during energetic flares is included. None of these works have studied the generation of possible false-positive biosignatures by time-dependent UV- flare radiation in atmospheres similar to prebiotic Earth. 
The objective of the present work is to shed light on possible false positives for O\textsubscript{2} and O\textsubscript{3} as biosignatures produced by stellar UV flux for Earth-like planets orbiting M-dwarfs with an atmosphere similar to Earth’s during the Hadean Eon (4.6-4.0 Ga), before the advent of life on Earth. This is relevant to understanding the possible scenarios that we might encounter with instruments that will be able to characterize potentially habitable planets with missions such as JWST, Large Interferometer For Exoplanets \citep{quanz_atmospheric_2021}, or an IR-visible-UV surveyor mission with the capability of characterizing a small number of nearby M-dwarf systems in reflected light \citep{stark_exoearth_2019, the_luvoir_team_luvoir_2019, NAP26141}. Note that throughout this work we assume that O atoms are primarily sourced from photolysis of O-bearing species such as CO\textsubscript{2} and H\textsubscript{2}O, i.e., there is no abundant free O\textsubscript{2}. The one-flare case is needed before exploring the effect of a series of flares to test the expected behavior of O\textsubscript{2} and O\textsubscript{3}, which should increase their abundances from enhanced CO\textsubscript{2} photolysis, but we found more complex trends that have not been reported before. The exploration of the impact of flares on abiotic O\textsubscript{2} atmospheres where significant O\textsubscript{2} is left behind due to massive hydrogen escape and an overwhelming of surface sinks is reserved for future work. 

\section{Oxygen atmospheric chemistry} \label{sec:O_atm_chem}

We focus on the photochemical mechanism to produce abiotic O\textsubscript{2} and O\textsubscript{3} on planets with CO\textsubscript{2}--N\textsubscript{2} atmospheres. Oxygen atoms are produced by the photolysis of CO\textsubscript{2}, which is recombined via reactions with HO\textsubscript{x} compounds that are produced by the photolysis of water, because the reaction
\begin{equation}\label{reaction:CO+O}
    \mathrm{CO} +\mathrm{O} + \mathrm{M} \longrightarrow \mathrm{CO}_2 + \mathrm{M}
\end{equation}
where ``M'' is any molecule in the atmosphere, is spin forbidden \citep{yung_photochemistry_1999}. The reactions that participate on the production and destruction of O, O\textsubscript{2}, and O\textsubscript{3} (O\textsubscript{x}) in these atmospheres are as follows:
\begin{enumerate}
    \item CO\textsubscript{2} photolysis:
    \begin{subequations}\label{reaction:CO2photo}
    \begin{equation}\label{reaction:CO2photoO}
        \mathrm{CO}_2 + \mathrm{h}\nu \longrightarrow \mathrm{CO} + \mathrm{O}\quad (\lambda \lesssim 200\,\mathrm{nm})
    \end{equation}
    \begin{equation}\label{reaction:CO2photoO1D}
        \mathrm{CO}_2 + \mathrm{h}\nu \longrightarrow \mathrm{CO} + \mathrm{O}(^{1}\mathrm{D})\quad (\lambda < 167\,\mathrm{nm})
    \end{equation}
    \end{subequations}
    Where O(\textsuperscript{1}D) is an oxygen radical that is highly reactive and thus short lived, and as such can return to a ground state via collision O(1D) + M $\rightarrow$ O + M or react with other chemical species. 
    
    \item O\textsubscript{2} formation:
    \begin{equation}\label{reaction:O+O}
        \mathrm{O} + \mathrm{O} + \mathrm{M} \longrightarrow \mathrm{O}_2 + \mathrm{M}
    \end{equation}
    Thus, the net result of the CO\textsubscript{2} photolysis is 2CO\textsubscript{2} $\rightarrow$ 2CO + O\textsubscript{2}.
    
    \item The Chapman cycle:
    \begin{subequations}\label{reaction:O2photo}
    \begin{equation}\label{reaction:O2photoO}
        \mathrm{O}_2 + \mathrm{h}\nu \rightarrow \mathrm{O} + \mathrm{O} \quad(175\,\mathrm{nm} < \lambda < 242\, \mathrm{nm})
    \end{equation}
     \begin{equation}\label{reaction:O2photoO1D}
        \mathrm{O}_2 + \mathrm{h}\nu \longrightarrow \mathrm{O} + \mathrm{O}(^{1}\mathrm{D})\quad (\lambda < 175\, \mathrm{nm})
    \end{equation}
    \end{subequations}   
    \begin{equation}\label{reaction:O+O2}
        \mathrm{O} + \mathrm{O}_2 + \mathrm{M} \longrightarrow \mathrm{O}_3 + \mathrm{M}
    \end{equation}
    \begin{subequations}\label{reaction:O3photo}
    \begin{equation}\label{reaction:O3photoO}
        \mathrm{O}_3 + \mathrm{h}\nu \longrightarrow \mathrm{O}_2 + \mathrm{O}\quad (310\, \mathrm{nm} < \lambda < 1140\, \mathrm{nm})
    \end{equation}
    \begin{equation}\label{reaction:O3photoO1D}
        \mathrm{O}_3 + \mathrm{h}\nu \longrightarrow \mathrm{O}_2 + \mathrm{O}(^{1}\mathrm{D})\quad (\lambda < 310\, \mathrm{nm})
    \end{equation}
    \end{subequations}
    \begin{equation}\label{reaction:O+O3}
        \mathrm{O} + \mathrm{O}_3 \longrightarrow 2\mathrm{O}_2
    \end{equation}  
   Reactions (\ref{reaction:O3photo}) do not destroy ozone because the reaction (\ref{reaction:O+O2}) quickly recombines this molecule. In this cycle, ozone is lost via reaction (\ref{reaction:O+O3}).
   
    \item The HO\textsubscript{x} catalytic cycles that start with the generation of H and OH via the reactions
    \begin{equation}\label{reaction:H2Ophoto}
        \mathrm{H}_{2}\mathrm{O} + h\nu \longrightarrow \mathrm{OH} + \mathrm{H}
    \end{equation}
    \begin{equation}\label{reaction:H2O+O1D}
        \mathrm{H}_{2}\mathrm{O} + \mathrm{O}(^{1}\mathrm{D}) \longrightarrow \mathrm{OH} + \mathrm{OH}
    \end{equation}

    Molecular hydrogen is also a source of HO\textsubscript{x} via the reactions
\begin{equation}\label{reaction:H2+O1D}
        \mathrm{H}_2 + \mathrm{O}(^{1}\mathrm{D}) \longrightarrow \mathrm{OH} + \mathrm{H}
    \end{equation}
\begin{equation}\label{reaction:H2+O}
        \mathrm{H}_2 + \mathrm{O} \longrightarrow \mathrm{OH} + \mathrm{H}
    \end{equation}
   
    After OH and H have been formed O\textsubscript{2} may be created via the catalytic cycles
    \begin{subequations}\label{cycle:OHcycle}
    \begin{equation}\label{reaction:OH+O3}
        \mathrm{OH} + \mathrm{O}_3 \longrightarrow \mathrm{HO}_2 + \mathrm{O}_2 
    \end{equation}
    \begin{equation}\label{reaction:HO2+O}
        \mathrm{HO}_2 + \mathrm{O} \longrightarrow \mathrm{OH} +\mathrm{O}_2
    \end{equation}
    \end{subequations}
    and
    \begin{subequations}\label{cycle:Hcycle}
    \begin{equation}\label{reaction:H+O3}
        \mathrm{H} + \mathrm{O}_3 \longrightarrow \mathrm{OH} + \mathrm{O}_2
    \end{equation}
    \begin{equation}\label{reaction:OH+O}
        \mathrm{OH} + \mathrm{O} \longrightarrow \mathrm{H} + \mathrm{O}_2
    \end{equation}
    \end{subequations}
    The net result of cycles \ref{cycle:OHcycle} and  \ref{cycle:Hcycle} is O + O\textsubscript{3} $\rightarrow$ 2O\textsubscript{2}.
    The other cycle that creates O\textsubscript{2} from ozone involves reactions 
    \begin{equation}
     \mathrm{OH} + \mathrm{O}_3 \longrightarrow \mathrm{HO}_2 + \mathrm{O}_2
     \tag{\ref{reaction:OH+O3}}
     \end{equation}
    \begin{equation}\label{reaction:HO2+O3}
        \mathrm{HO}_2 + \mathrm{O}_3 \longrightarrow \mathrm{OH} + 2\mathrm{O}_2
    \end{equation}    
    The net result of cycle (\ref{reaction:HO2+O3}) is 2O\textsubscript{3} $\rightarrow$ 3O\textsubscript{2}.
    
    \item The NO\textsubscript{x} catalytic cycle that destroys O\textsubscript{3}:
    \begin{subequations}\label{cycle:NOcycle}
    \begin{equation}\label{reaction:NO+O3}
        \mathrm{NO} + \mathrm{O}_3 \longrightarrow \mathrm{NO}_2 + \mathrm{O}_2
    \end{equation}
    \begin{equation}\label{reaction:NO2+O}
        \mathrm{NO}_2 + \mathrm{O} \longrightarrow \mathrm{NO} + \mathrm{O}_2
    \end{equation}
    \end{subequations}
    The net result here is O\textsubscript{3} + O $\rightarrow$ 2O\textsubscript{2} (cycle \ref{cycle:NOcycle}). There are other catalytic cycles that can lead to the destruction of O\textsubscript{3}, including those involving Cl, Br, and others \citep[e.g.][]{anderson_kinetics_1989} but these are less important than the above reactions for the case of an anoxic Earth-like terrestrial planet with moderate volcanism. 
    
    \item The CO\textsubscript{2} photolysis followed by HO\textsubscript{x} reactions:
     \begin{equation}
     2(\mathrm{CO}_2 + \mathrm{h}\nu \longrightarrow \mathrm{CO} + \mathrm{O})
     \tag{\ref{reaction:CO2photoO}}
     \end{equation}
    \begin{equation}
     \mathrm{OH} + \mathrm{O} \longrightarrow \mathrm{H} + \mathrm{O}_2
     \tag{\ref{reaction:OH+O}}
     \end{equation}
      \begin{equation}\label{reaction:H+O2}
        \mathrm{H} + \mathrm{O}_2 + \mathrm{M} \longrightarrow \mathrm{HO}_2 + \mathrm{M}
    \end{equation}
    \begin{equation}
     \mathrm{HO}_2 + \mathrm{O} \longrightarrow \mathrm{OH} + \mathrm{O}_2
     \tag{\ref{reaction:HO2+O}}
     \end{equation}
     The net result for these reactions is 2CO\textsubscript{2}~$\rightarrow$~2CO~+~O\textsubscript{2} \citep{yung_photochemistry_1999}. Notice that for low CO\textsubscript{2} atmospheres, like present Earth, reaction \ref{reaction:CO2photoO} is not the main source of free oxygen, thus this HO\textsubscript{x} catalytic cycle results in 2O~$\rightarrow$~O\textsubscript{2} \citep{kozakis_is_2022}.
    
    \item The recombination of CO\textsubscript{2} mediated by HO\textsubscript{x} reactions:
    several cycles contribute to the recombination of CO\textsubscript{2}, which results in O\textsubscript{2} production or loss, or O\textsubscript{3} loss:  
     \begin{equation}
        2 ( \mathrm{H} + \mathrm{O}_2 + \mathrm{M} \longrightarrow \mathrm{HO}_2 + \mathrm{M} )
     \tag{\ref{reaction:H+O2}}
    \end{equation}
     \begin{equation}\label{reaction:2HO2}
        \mathrm{HO}_2 + \mathrm{HO}_2 \longrightarrow \mathrm{H}_2\mathrm{O}_2 + \mathrm{O}_2
    \end{equation}
    \begin{equation}\label{reaction:H2O2photo}
       \mathrm{H}_2\mathrm{O}_2 + h\nu \longrightarrow \mathrm{OH} + \mathrm{OH}
    \end{equation}
    \begin{equation}\label{reaction:OH+CO}
       2(\mathrm{OH} + \mathrm{CO} \longrightarrow \mathrm{CO}_2 + \mathrm{H})
    \end{equation}
The net result of these reactions is 2CO + O\textsubscript{2}~$\rightarrow$~2CO\textsubscript{2}. When O\textsubscript{3} becomes more abundant then, the former pathway shifts to
    \begin{equation}
     \mathrm{H} + \mathrm{O}_2 + \mathrm{M} \longrightarrow \mathrm{HO}_2 + \mathrm{M}
     \tag{\ref{reaction:H+O2}}
    \end{equation}
    \begin{equation}
        \mathrm{HO}_2 + \mathrm{O}_3 \longrightarrow \mathrm{OH} + 2\mathrm{O}_2
    \tag{\ref{reaction:HO2+O3}}
    \end{equation} 
     \begin{equation}
       \mathrm{OH} + \mathrm{CO} \longrightarrow \mathrm{CO}_2 + \mathrm{H}
        \tag{\ref{reaction:OH+CO}}
    \end{equation}
    The net result for this set of reactions is CO + O\textsubscript{3}~$\rightarrow$~CO\textsubscript{2} + O\textsubscript{2} \citep{grenfell2013,gao_stability_2015}.
    The last three chemical paths depend on the abundance of HO\textsubscript{2} and H\textsubscript{2}O\textsubscript{2} which work as a storage of OH that is released upon their photolysis \citep{gao_stability_2015,harman_abiotic_2015}.
\end{enumerate}

\section{Methods}\label{sec:methods}
\subsection{Photochemical model}
We used the photochemical model component of the Atmos package \citep{arney_pale_2016, lincowski_evolved_2018} as our base photochemical code, which was further adapted to predict time-dependent atmospheric compositions during flare events as described below. We adopted an Archean Earth template reaction network with 74 chemical species and 392 reactions. The code includes vertical mixing and considers altitudes from 0 km (the surface) to 100 km. We incorporated the newest H\textsubscript{2}O and CO\textsubscript{2} cross sections according to the recommendations of \citet{ranjan_photochemistry_2020} and \citet{broussard_impact_2024}. Our code differs from some earlier branches of the same model by allowing the CO\textsubscript{2} mixing ratio to be fixed only at the surface so that it can change along the atmosphere during the flare. The model includes NO\textsubscript{x} production by lightning \citep{harman_abiotic_2018}. The Atmos code is widely used to predict the atmospheric composition of terrestrial exoplanets with secondary atmospheres \citep{lustig-yaeger_detectability_2019, peacock_accurate_2022, teal_effects_2022}.

Due to the particular needs of working with the changing spectra of a star during a flare, the code was further modified to be capable of tracking the atmospheric evolution over the short timescales of a flare \citep{segura_effect_2010, tilley_modeling_2019}. Following the scheme outlined by \citet{segura_effect_2010} a control layer was programmed on top of an updated version of the Atmos photochemical model \citep{arney_pale_2016}; this layer informs the main routine of the stellar status, keeps control of time, and hands the stellar flux and atmospheric composition as required. Another time dependent version of Atmos exists developed by \citet{wogan_photochempy_2022} (Photochempy) that uses a the CVODEBDF Ordinary Differential Equations solver from Sundials Computing, this solver is an implementation of the backward differential formulas (BDF), instead of the Backward Euler method used in Atmos.

The status of a stellar spectrum is divided into three stages: pre-, post-, and during flare. Pre-flare gives free reign to the main routine, allowing the atmosphere to reach a stationary state; the atmosphere of this pre-flare stage is passed over to the next stage. During the flare, time is controlled, giving pauses as the flare probing requires, passing on the atmosphere to the next step while changing the stellar flux incident in the planet. Finally, during the post-flare stage, the main routine is given freedom to run until the atmosphere reaches a stationary state while regularly monitoring the state of the atmosphere; this stage begins with the atmosphere as it was at the end of the second stage.
\subsection{Flare input}

\begin{figure}[ht!]
\includegraphics[width=\textwidth]{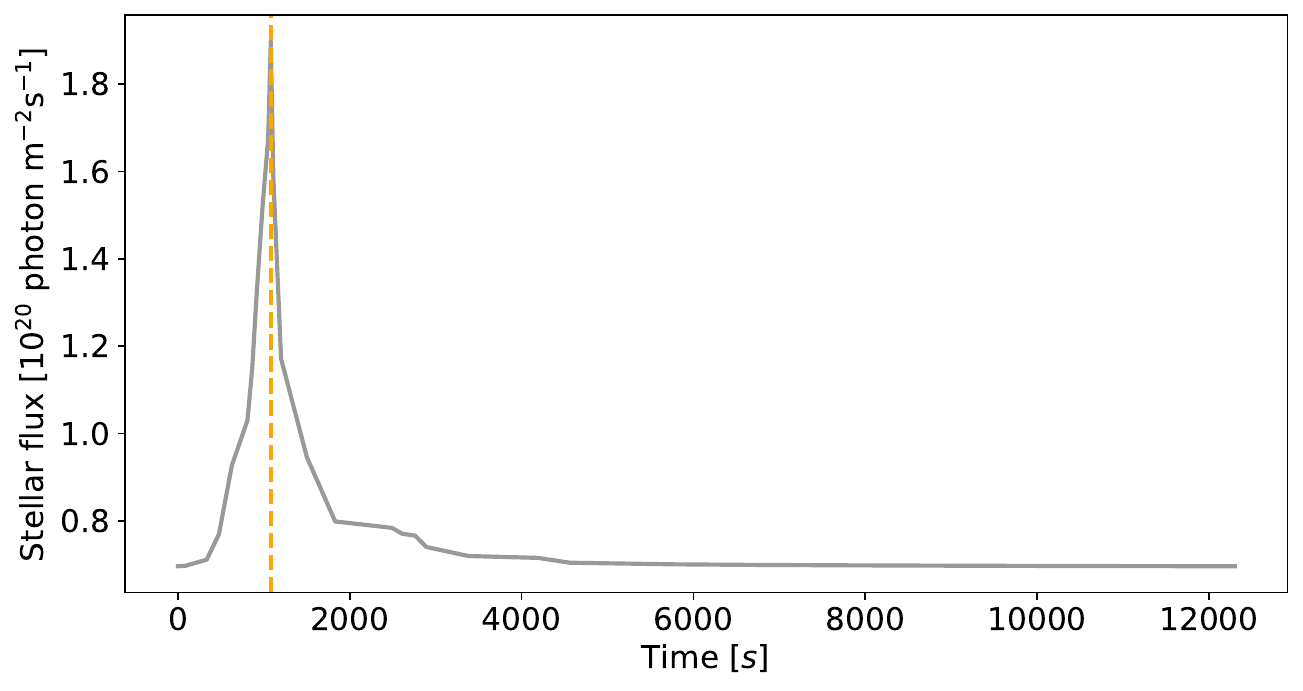}
\caption{Light curve for AD Leo during the flare. The orange line highlights the maximal flux emission during the flare at 1080 s.  
\label{fig1:lightcurve}}
\end{figure}

\begin{figure}[ht!]
\includegraphics[width=\textwidth]{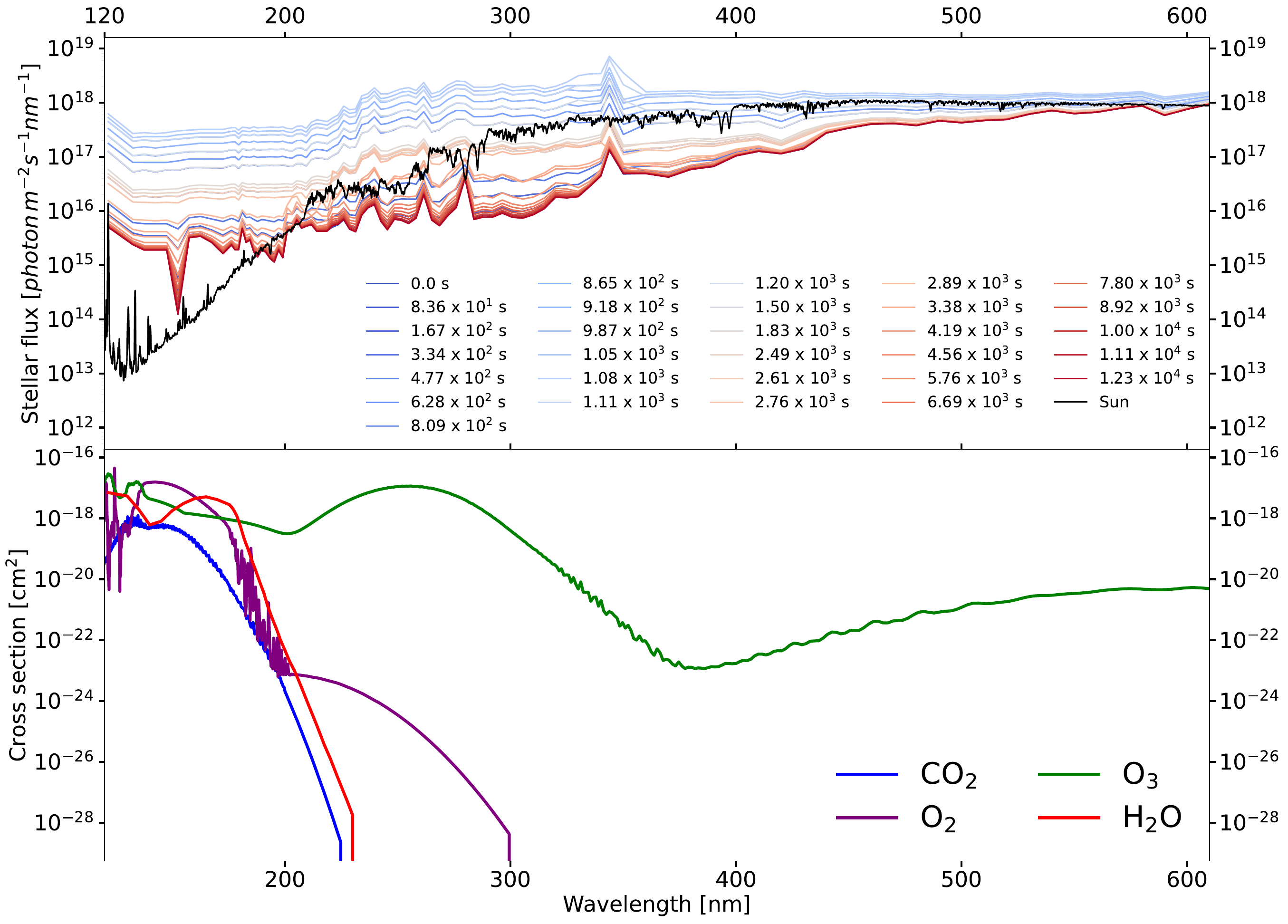}
\caption{Upper panel: Flare spectra used for the simulations. The solar flux at the top of the Earth’s atmosphere is shown in black. Lower panel: Absorption cross sections of H\textsubscript{2}O, CO\textsubscript{2}, o\textsubscript{2} and O\textsubscript{3}.  
\label{fig2:spectra-crosssections}}
\end{figure}

The light curve is based on the empirical results from \citet{davenport_kepler_2014} and it is sampled in 30 points \citep{tilley_modeling_2019}; see Figure \ref{fig1:lightcurve}. The flare spectrum evolution used the fluxes as described in \citet{tilley_modeling_2019}; see Figure \ref{fig2:spectra-crosssections} (top). The total energy of the flare is 10\textsuperscript{34} erg. The flare used in this work is ultimately based on the Great AD Leo flare of 1985 and reported by \citet{hawley_great_1991}. It is the flare often used in the literature because it is the only one that was measured in the NUV (200-350 nm) and far-UV (FUV; 100-200 nm). \citet{tilley_modeling_2019} scaled this flare to produce spectra of less energetic flares, but here we decided to use the original high-energy flare because it would illustrate an extreme case of O\textsubscript{2}/O\textsubscript{3} abiotic production.

A second high-energy flare was also used in simulations, following the same light-curve-profile as before, but scaling the total energy by 2 orders of magnitude, to a total energy of the flare of 10\textsuperscript{36} erg. This simulations are intended as end-members to view the effect of a super flare, even if less common than 10\textsuperscript{34} erg flares, on this type of planets.

\subsection{Characteristics of the simulated atmospheres}

\begin{deluxetable}{llll}
\tabletypesize{\scriptsize}
\tablewidth{0pt} 
%\tablenum{1}
\tablecaption{Boundary Conditions for the Simulated Atmospheres.}\label{tab1:boundary}
\tablehead{
\colhead{Compound} & \colhead{Boundary Conditions} & \colhead{Prebiotic Earth-like} & \colhead{Low Hydrogen} 
} 
\startdata
O\textsubscript{2}&Constant deposition velocity&0.0 cm s\textsuperscript{-1}&0.0 cm s\textsuperscript{-1}\\
H\textsubscript{2}&Constant deposition velocity&0.0 cm s\textsuperscript{-1}&\nodata\\
&Vertically distributed upward flux&$3.0\times10^{10}$ molecules cm\textsuperscript{-2} s\textsuperscript{-1}&\nodata\\
&Constant mixing ratio&\nodata&$1.0\times10^{-10}$\\
CO&Constant deposition velocity&1.0 $\times$ 10\textsuperscript{-8} cm s\textsuperscript{-1}&1.0 $\times$ 10\textsuperscript{-8} cm s\textsuperscript{-1}\\
CH\textsubscript{4}&Constant surface flux&$6.8\times10^{8}$ molecules cm\textsuperscript{-2} s\textsuperscript{-1}&\nodata\\
&Constant mixing ratio&\nodata&1.0$\times10^{-8}$\\
H\textsubscript{2}S&Constant deposition velocity&$1.5\times10^{-2}$ cm s\textsuperscript{-1}&\nodata\\
&Vertically distributed upward flux&$3.5\times10^{8}$ molecules cm\textsuperscript{-2} s\textsuperscript{-1}&\nodata\\
&Constant mixing ratio&\nodata&$1.0\times10^{-10}$
\enddata
\end{deluxetable}

The composition of the simulated atmospheres was designed to be similar to the prebiotic Earth because on an Earth-like planet without life the atmosphere would be the product of mantle degassing through volcanism, which should result in large quantities of N\textsubscript{2}, CO\textsubscript{2}, and water with minor quantities of H\textsubscript{2}, CO, and CH\textsubscript{4} \citep{tian_hydrogen-rich_2005, catling_comment_2006, zahnle_earths_2010}. The simulated atmospheres had 3\%, 10\%, 30\%, 60\%, and 80\% CO\textsubscript{2} with a surface pressure of 1 bar (N\textsubscript{2} was used as a filler gas). The planets were located at 1 au equivalent distance (i.e. they received the same integrated flux as present Earth from the Sun). The boundary conditions for the most relevant atmospheric compounds are described in Table~\ref{tab1:boundary}.

\citeauthor{harman_abiotic_2015} (\citeyear{harman_abiotic_2015, harman_abiotic_2018}) demonstrated that the selection of boundary conditions is key for the false-positives problem. Following their work, we selected a deposition velocity for CO of 1 $\times$ 10$^{-8}$ cm s\textsuperscript{-1}, consistent with the abiotic formation of formate \citep{harman_abiotic_2015}, and zero for O\textsubscript{2} to maximize the accumulation of O\textsubscript{x} in the atmosphere.

Atmospheres are given a different treatment based on their CO\textsubscript{2} content, considering that a 60\% mixing ratio is enough to think of it as the dominant background gas. Water rainout and condensation are treated equally in high- and low-CO\textsubscript{2} atmospheres, the same as saturation pressure. Molecular diffusion of H and H\textsubscript{2} in the upper parts of the atmosphere in high-CO\textsubscript{2} atmospheres account for the CO\textsubscript{2} becoming the dominant background gas. For all CO\textsubscript{2} abundances we simulated one atmosphere with a volcanic flux similar to those typically assumed for early Earth (abiotic fluxes only, ``high H\textsubscript{2};" \citeauthor{Holland2002} \citeyear{Holland2002}, \citeauthor{Kharecha2005} \citeyear{Kharecha2005}, \citeauthor{guzman-marmolejo_abiotic_2013} \citeyear{guzman-marmolejo_abiotic_2013}, \citeauthor{domagal-goldman_abiotic_2014} \citeyear{domagal-goldman_abiotic_2014}) and one with low hydrogen sources and availability (``low H\textsubscript{2},'' see Table~\ref{tab1:boundary}). The low-H\textsubscript{2} end-member features a low fixed H\textsubscript{2} mixing ratio, which could occur in the presence of abundant surface oxidants.

We simulated two additional atmospheres to test the maximum abiotic production of oxygen and ozone with no hydrogen sources, given that hydrogen-bearing compounds are a sink of oxygen compounds \citep{segura_abiotic_2007}. Thus, we have an 80\% CO\textsubscript{2} ``desiccated atmosphere'' where relative humidity is lowered down to levels samller than the desert of Atacama ($1.0\times10^{-3}$ in Atacama vs $1.0\times10^{-8}$ used;  H\textsubscript{2}O surface mixing ratio of $~1.0\times10^{-10}$), as to ensure low H\textsubscript{2}O levels without decreasing H\textsubscript{2}O so low as to fall into a CO runaway-production scenario, as found by \citet{gao_stability_2015} and \citet{zahnle_photochemical_2008}. This atmosphere was studied with and without the effect of lightning, to see the effect NO\textsubscript{x} species have on the production of ozone. Unless specified, the text will refer to the lightning-less scenario as the ``desiccated atmosphere,'' as the lack of water would prevent cloud formation and therefore lightning occurrence.  We emphasize that these atmospheric scenarios are not necessarily meant to be prospective predictions of the atmospheric composition of any particular planet, but were chosen as end-members to explore the relative impact of a single UV flare on atmospheres with different preexisting oxidation states.

\subsection{Planetary spectra}\label{subsec:Pspectra}

We computed synthetic reflectance, emission, and transmission spectra for select cases with the Spectral Mapping Atmospheric Radiative Transfer (SMART) model \citep{meadows_ground-based_1996, crisp_absorption_1997}. SMART is a line-by-line, fully multiple scattering radiative-transfer code capable of generating synthetic spectral data from FUV to far-IR wavelengths. SMART has been well validated by observations of Venus \citep{arney_spatially_2014},  Earth \citep{robinson_earth_2011},  Mars \citep{tinetti_disk-averaged_2005}, and Titan \citep{robinson_titan_2014}, and it has been used extensively for predicting the spectral observables of exoplanet atmospheres \citep{charnay_3d_2015, barnes_habitability_2018, mandt_trappist-1h_2022}. For simulated transit transmission observations, we use the ray-tracing model of \citep{robinson_theory_2017}, which includes the effects of refraction. We calculated molecular absorption coefficients using the HITRAN 2016 database \citep{gordon_hitran2016_2017}. We assume planetary parameters consistent with the photochemical simulations when constructing the spectral simulations. To calculate transit depth, we assume a host-star radius equivalent to that of TRAPPIST-1 \citep[$R$=0.121 $R$\textsubscript{$\earth$},][]{grootel_stellar_2018}. We note that the TRAPPIST-1 radius is not self-consistent with our stellar spectrum, but we choose the radius of an M8V star to illustrate the maximum size of abiotic O\textsubscript{2}/O\textsubscript{3} features induced by a flare and because near-term observations of terrestrial exoplanetary atmospheres are more likely for the TRAPPIST-1 system than any other system. Reflected light observations are degraded to a spectral resolving power of $R=400$ while transmission observations are degraded to $R=200$. 

\section{Results} \label{sec:results}

\subsection{Oxygen and ozone abundances}

\subsubsection{Steady-state results}

\begin{deluxetable}{lcccc}
\tabletypesize{\scriptsize}
\tablewidth{0pt} 
%\tablenum{1}
\tablecaption{Pre-flare oxygen and ozone columns densities. Present Earth values are $4.65\times10^{24}$ cm\textsuperscript{-2} for O\textsubscript{2} and $8.61\times10^{18}$ cm\textsuperscript{-2} for O\textsubscript{3}.\label{tab2:preflarecols}}
\tablehead{
&\multicolumn{2}{c}{High H\textsubscript{2}}&\multicolumn{2}{c}{Low H\textsubscript{2}}\\
\colhead{CO\textsubscript{2} mixing ratio} & \colhead{O\textsubscript{2} column depth (cm\textsuperscript{-2})}& \colhead{O\textsubscript{3} column depth (cm\textsuperscript{-2})} & \colhead{O\textsubscript{2} column depth (cm\textsuperscript{-2})}&\colhead{O\textsubscript{3} column depth (cm\textsuperscript{-2})}
} 
\startdata
0.03&$1.02\times10^{20}$&$2.33\times10^{15}$&$1.18\times10^{20}$&$5.95\times10^{15}$\\
0.1&$1.34\times10^{20}$&$2.35\times10^{15}$&$1.51\times10^{20}$&$6.11\times10^{15}$\\
0.3&$1.55\times10^{20}$&$2.14\times10^{15}$&$1.95\times10^{20}$&$7.09\times10^{15}$\\
0.6&$1.59\times10^{20}$&$3.57\times10^{15}$&$2.02\times10^{20}$&$1.15\times10^{16}$\\
0.8&$1.46\times10^{20}$&$2.55\times10^{15}$&$1.85\times10^{20}$&$9.85\times10^{15}$\\
0.8 desiccated lightning&\nodata&\nodata&$2.79\times10^{20}$&$2.40\times10^{17}$\\
0.8 desiccated lightning-less&\nodata&\nodata&$4.28\times10^{24}$&$1.28\times10^{19}$\\
\enddata
\end{deluxetable}

\begin{figure}[ht!]
\includegraphics[width=\textwidth]{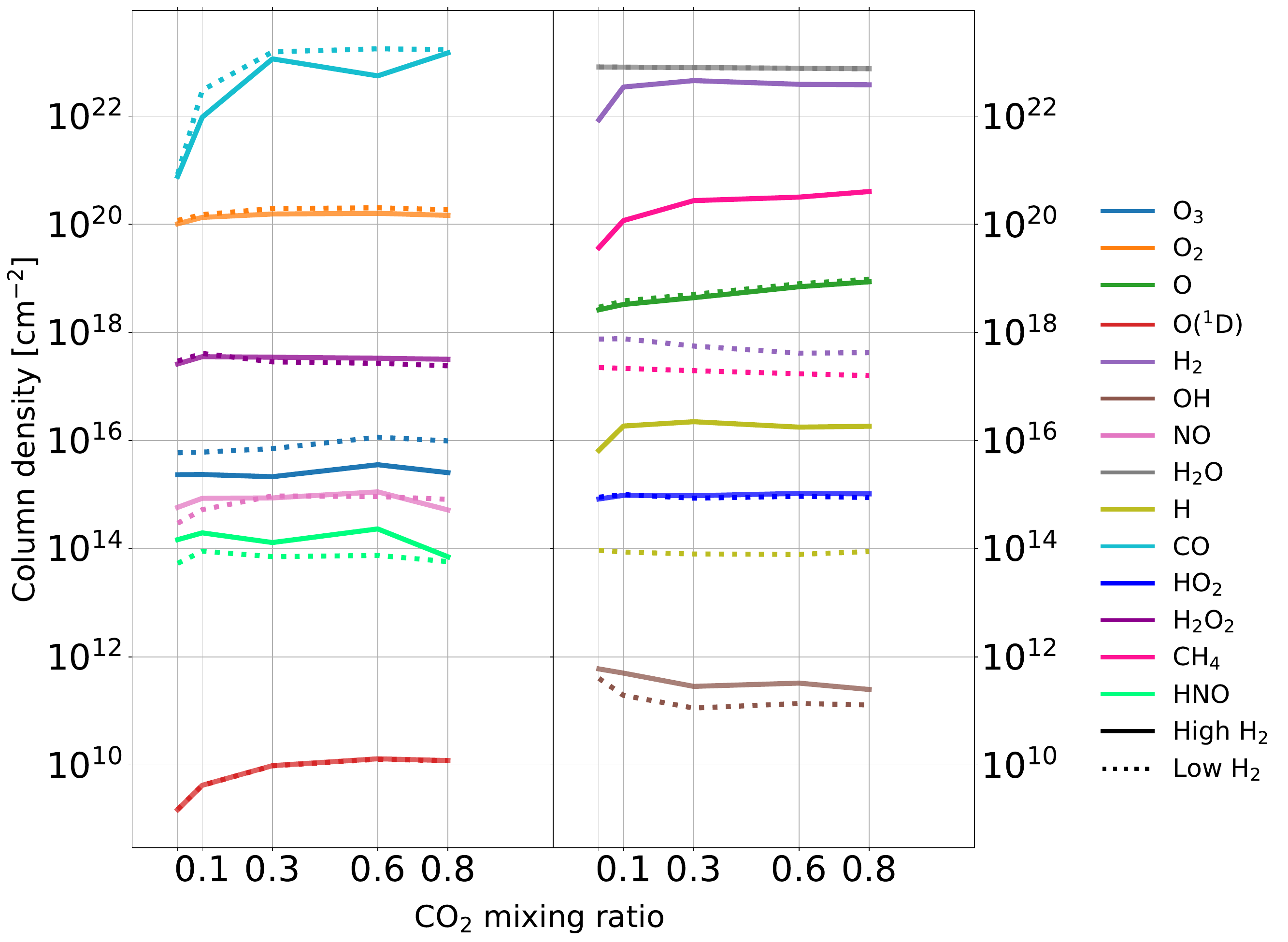}
\caption{Relevant column densities for the ozone and molecular oxygen abundances. The chemical species are divided in two panels for clarity. High H\textsubscript{2} atmospheres in solid lines, low H\textsubscript{2} atmospheres in dotted lines. 
\label{fig:columndepths-steady}}
\end{figure}

The initial (pre-flare) column densities for O\textsubscript{2} and O\textsubscript{3} for our simulated atmospheres are listed in Table~\ref{tab2:preflarecols}. CO\textsubscript{2} photolysis is the source of O and O(\textsuperscript{1}D) that initiates the production of O\textsubscript{2} and O\textsubscript{3}, after that the Chapman cycle and the catalytic reactions with NO\textsubscript{x} and HO\textsubscript{x} determine the O\textsubscript{2} and O\textsubscript{3} abundances. Figure~\ref{fig:columndepths-steady} presents the initial abundances for O\textsubscript{2}, O\textsubscript{3}, H\textsubscript{2}, O,  O($^1$D), OH, and NO, the last two as representatives of the HO\textsubscript{x} and NO\textsubscript{x} compounds, respectively.

\paragraph{Water, H\textsubscript{2}, and HO\textsubscript{x}} The water abundance in the troposphere is set up using a fixed humidity that depends on the temperature. Because the troposphere is convective, the temperature is defined by the adiabatic lapse rate. There is abundant water in the troposphere ($\sim$1\%) for the non-desiccated cases, supported by an assumed equilibrium with a surface ocean reservoir. The cold trap at the tropopause \textquotedblleft leaks" some water to the stratosphere at around 10 km. At higher altitudes its mixing ratio is determined by photolysis (reaction (\ref{reaction:H2Ophoto})) and the following recombination reactions:

\begin{equation}\label{reaction:H2+OH}
\mathrm{H}_2 + \mathrm{OH} \longrightarrow \mathrm{H}_2\mathrm{O} + \mathrm{H}
\end{equation}
\begin{equation}\label{reaction:H+HO2}
\mathrm{H} + \mathrm{HO}_2 \longrightarrow \mathrm{H}_2\mathrm{O} + \mathrm{O}
\end{equation}
\begin{equation}\label{reaction:OH+HO2}
\mathrm{OH} + \mathrm{HO}_2 \longrightarrow \mathrm{H}_2\mathrm{O} + \mathrm{O}_2
\end{equation}
\begin{equation}\label{reaction:H2O2+OH}
\mathrm{H}_2 \mathrm{O}_2 + \mathrm{OH} \longrightarrow \mathrm{H}_2\mathrm{O} + \mathrm{HO}_2
\end{equation}

\begin{figure}[ht!]
\includegraphics[width=\textwidth]{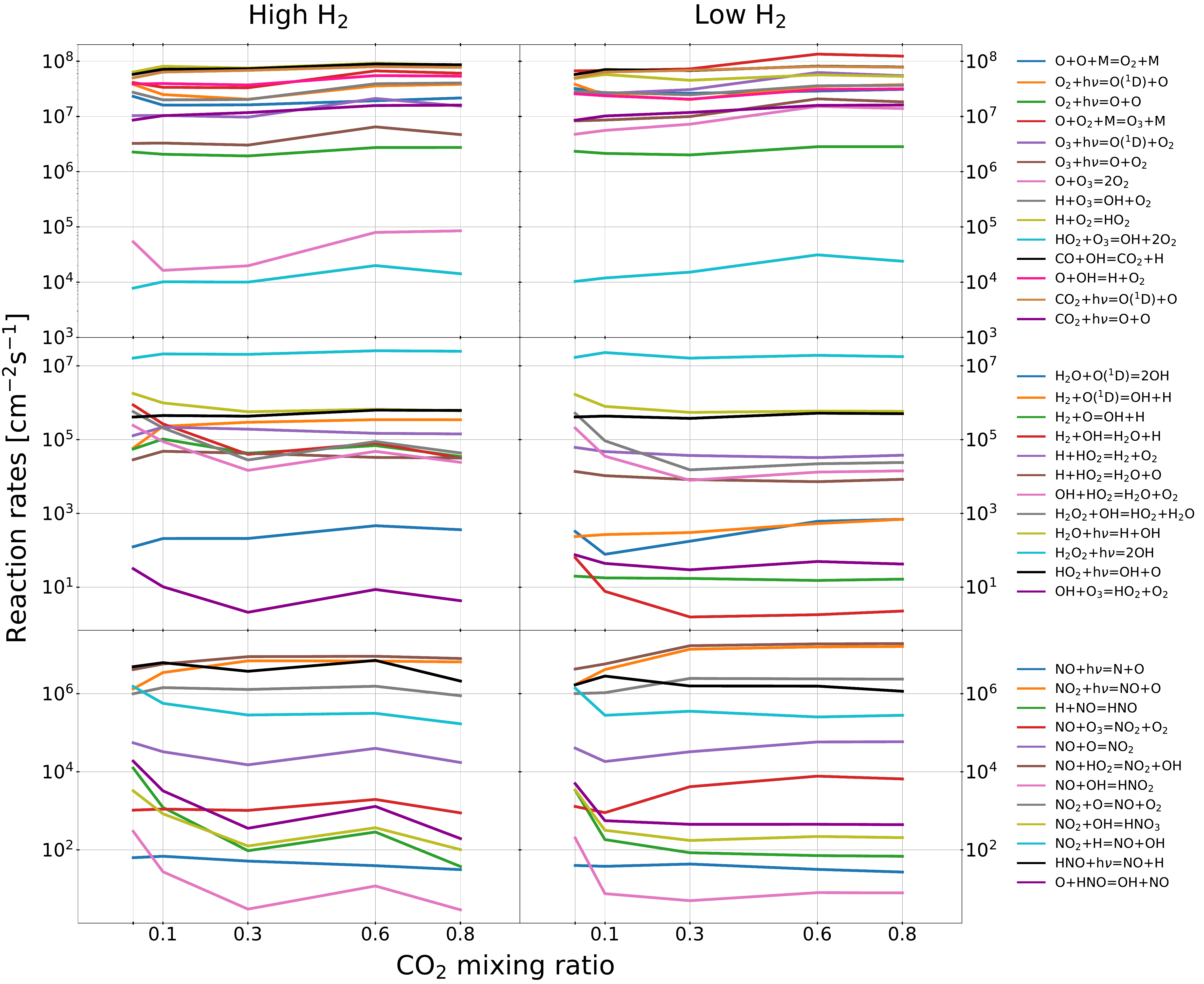}
\caption{Relevant reaction rates for O\textsubscript{x} compounds (upper panel),
hydrogen-bearing chemical species (middle panel), and NO\textsubscript{x} (lower panel), for atmospheres with high-H sources (left column) and with suppressed-H sources (right panels).
\label{fig:reactionrates-steady}}
\end{figure}

On modern Earth, O\textsubscript{2} protects water from photolysis; if O\textsubscript{2} decreases but has a constant mixing ratio throughout the atmosphere, then there is less water in the stratosphere \citep{kozakis_is_2022}. In CO\textsubscript{2}-N\textsubscript{2}-dominated atmospheres, water exhibits a different pattern, decreasing by around 6\% with increasing CO\textsubscript{2}. In these atmospheres, CO\textsubscript{2} acts as a weak UV shield for water, then, as CO\textsubscript{2} increases, water photolysis decreases only by a few percent ($\sim$10\%, see Fig. \ref{fig:reactionrates-steady}, middle panels, olive lines). The main reaction that recombines water (reaction (\ref{reaction:H2+OH}), see Fig. \ref{fig:reactionrates-steady}, middle panels, red lines) becomes slower as the CO\textsubscript{2} rises, because there is slightly less OH available for high-CO\textsubscript{2} atmospheres. The atmospheres with low hydrogen have very similar water abundance to their counterparts with high hydrogen for a given CO\textsubscript{2} mixing ratio (the difference is in the fourth decimal and not noticeable in Figure~\ref{fig:columndepths-steady}), because most of the water is in the troposphere.

H\textsubscript{2} has a fixed flux at the surface for a subset of our atmospheres, the high-H\textsubscript{2} atmospheres. After its production, assuming the planet has volcanic activity, abundance depends on the reactions with oxygen atoms (reactions (\ref{reaction:H2+O1D}) and (\ref{reaction:H2+O}), fig. \ref{fig:reactionrates-steady}, yellow and green lines on middle panel) and the reaction that recombines water (reaction (\ref{reaction:H2+OH}), fig. \ref{fig:reactionrates-steady}, red line on middle panel). The latter is the dominant sink of H\textsubscript{2} in low-CO\textsubscript{2} atmospheres (Fig. \ref{fig:reactionrates-steady}, middle panel) but drastically drops as CO\textsubscript{2} increases and the OH abundance diminishes. Thus, the main sink of H\textsubscript{2} changes from reaction (\ref{reaction:H2+OH}) to the reactions (\ref{reaction:H2+O1D}) and (\ref{reaction:H2+O}), as O and O($^1$D) increase with higher CO\textsubscript{2}. The net effect is a maximum abundance of H\textsubscript{2} at 30\% CO\textsubscript{2} for the cases with high-H (Figure~\ref{fig:columndepths-steady}). The low-H\textsubscript{2} atmospheres have a fixed H\textsubscript{2} mixing ratio at the surface; without the volcanic flux, the sources for hydrogen-bearing species are the byproducts of water photolysis, which lowers with increasing CO\textsubscript{2} (as can be seen in figure \ref{fig:columndepths-steady}).

HO\textsubscript{x} participate in catalytic cycles that recombine CO\textsubscript{2} and destroy O\textsubscript{3} and O\textsubscript{2} (see Section \ref{sec:O_atm_chem}). OH decreases as CO\textsubscript{2} increases (Figure~\ref{fig:columndepths-steady}, brown line), as a direct result of less water photolysis (Fig. \ref{fig:reactionrates-steady}, middle panel, olive line). Atomic hydrogen exhibits a different trend, increasing with larger CO\textsubscript{2}. Even when the main source of OH and H (water photolysis, see fig. \ref{fig:reactionrates-steady} middle panel, olive line) has a symmetric production of both species, methane in the atmosphere quickly reacts with OH, accounting for the difference between both columns (fig. \ref{fig:columndepths-steady}, olive and brown lines).

\paragraph{NO\textsubscript{x}} In our model, NO\textsubscript{x} species are produced by lightning and depend on the abundance of N\textsubscript{2} and O\textsubscript{2}. The N\textsubscript{2} abundance goes from 97\% to 20\%, while O$_2$ has a minimum at the 3\% CO$_2$ atmospheres and reaches its maximum at the 60\% CO$_2$ atmospheres (Figure~\ref{fig:columndepths-steady}). As a result of the combination of N$_2$ and O$_2$ concentrations, the 3\% and 60\% CO$_2$ atmospheres have the lowest NO and the largest NO abundances, respectively. The NO abundance difference between the high- and low- H$_2$ atmospheres is due to the HO$_2$ that is the main NO sink (brown line in the lower panel of Figure~\ref{fig:reactionrates-steady}).

\paragraph{O\textsubscript{2} and O\textsubscript{3}} Molecular oxygen generally increases with CO\textsubscript{2} due to more atomic oxygen being produced via photolysis of carbon dioxide. This happens for all atmospheres, except for the 80\% CO\textsubscript{2} where opacity of the atmosphere gets to its maximum. As opacity reaches its maximum, water photolysis is diminished, this photolytical shielding suppresses the catalytic cycles involved in O\textsubscript{2} production. Within high-H\textsubscript{2} atmospheres the O\textsubscript{2} abundance responds mainly to the availability of O atoms, but also has sources in the catalytic cycles of HO\textsubscript{x} and NO\textsubscript{x}. Catalytic cycles \ref{cycle:OHcycle} and \ref{cycle:Hcycle} produce O\textsubscript{2}, however H+O\textsubscript{2} $\rightarrow$ HO\textsubscript{2} (reaction \ref{reaction:H+O2}) is a loss of oxygen, and a faster reaction, meaning the low-H\textsubscript{2} atmospheres  have more molecular oxygen than their high-H\textsubscript{2} counterparts.

For the both, low- and high-H cases, molecular oxygen increases with CO$_2$ abundance except for the 80\% atmosphere, where catalytic reactions that convert O$_3$ back to O$_2$ have a local minimum  (cycles \ref{cycle:Hcycle} and \ref{cycle:OHcycle}, Fig.~\ref{fig:reactionrates-steady}) due to the drop of OH that is consumed by methane. For high-H atmospheres, ozone has an opposite trend to H, its main sink (grey line in the upper left panel in Fig.~\ref{fig:reactionrates-steady}), this behavior is softened in the low-H atmospheres. For the low-H atmospheres ozone follows the same trend as molecular oxygen because its abundance is dominated by the third body reaction that creates ozone (red line in the upper right panel in Fig.~\ref{fig:reactionrates-steady}).

The inverse relationship between the abundances of hydrogen-bearing compounds and the abundances of O$_2$ and O$_3$ is further demonstrated by the increased column densities of the latter two gases in desiccated atmospheres (Table~\ref{tab2:preflarecols}). Similarly, an inverse relationship between O$_2$ and O$_3$ with NO$_x$ species can be seen, particularly as lightning is deactivated.
%The importance of hydrogen bearing compounds in the production of O\textsubscript{2} and O\textsubscript{3} is further evidenced by the increase in both abundances for the desiccated atmospheres (see table~\ref{tab2:preflarecols}), similarly for the nitrogen bearing compounds, as both abundances further increase as the nitrogen-bearing-compounds production by lighting is deactivated.

\subsubsection{Flare results}\label{subsubsection:flare-results}

\begin{deluxetable}{lcccc}
\tabletypesize{\scriptsize}
\tablewidth{0pt} 
%\tablenum{1}
\tablecaption{Maximum oxygen and ozone columns densities during the flare and recovery. Present Earth values are $4.65\times10^{24}$ cm\textsuperscript{-2} for O\textsubscript{2} and $8.61\times10^{18}$ cm\textsuperscript{-2} for O\textsubscript{3}.\label{tab3:maxflarecols}}
\tablehead{
&\multicolumn{2}{c}{with H\textsubscript{2}}&\multicolumn{2}{c}{without H\textsubscript{2}}\\
\colhead{CO\textsubscript{2} mixing ratio} & \colhead{O\textsubscript{2} column depth (cm\textsuperscript{-2})}& \colhead{O\textsubscript{3} column depth (cm\textsuperscript{-2})} & \colhead{O\textsubscript{2} column depth (cm\textsuperscript{-2})}&\colhead{O\textsubscript{3} column depth (cm\textsuperscript{-2})}
} 
\startdata
0.03&$1.02\times10^{20}$&$2.41\times10^{15}$&$1.18\times10^{20}$&$6.81\times10^{15}$\\
0.1&$1.34\times10^{20}$&$2.39\times10^{15}$&$1.51\times10^{20}$&$7.40\times10^{15}$\\
0.3&$1.55\times10^{20}$&$2.18\times10^{15}$&$1.95\times10^{20}$&$8.49\times10^{15}$\\
0.6&$1.60\times10^{20}$&$3.63\times10^{15}$&$2.02\times10^{20}$&$1.34\times10^{16}$\\
0.8&$1.46\times10^{20}$&$2.67\times10^{15}$&$1.86\times10^{20}$&$1.23\times10^{16}$\\
0.8 desiccated lightning&\nodata&\nodata&$2.79\times10^{20}$&$3.54\times10^{17}$\\
0.8 desiccated lightning-less&\nodata&\nodata&$4.28\times10^{24}$&$1.29\times10^{19}$\\
\enddata
\end{deluxetable}

\begin{figure}[hb!]
\includegraphics[width=\textwidth]{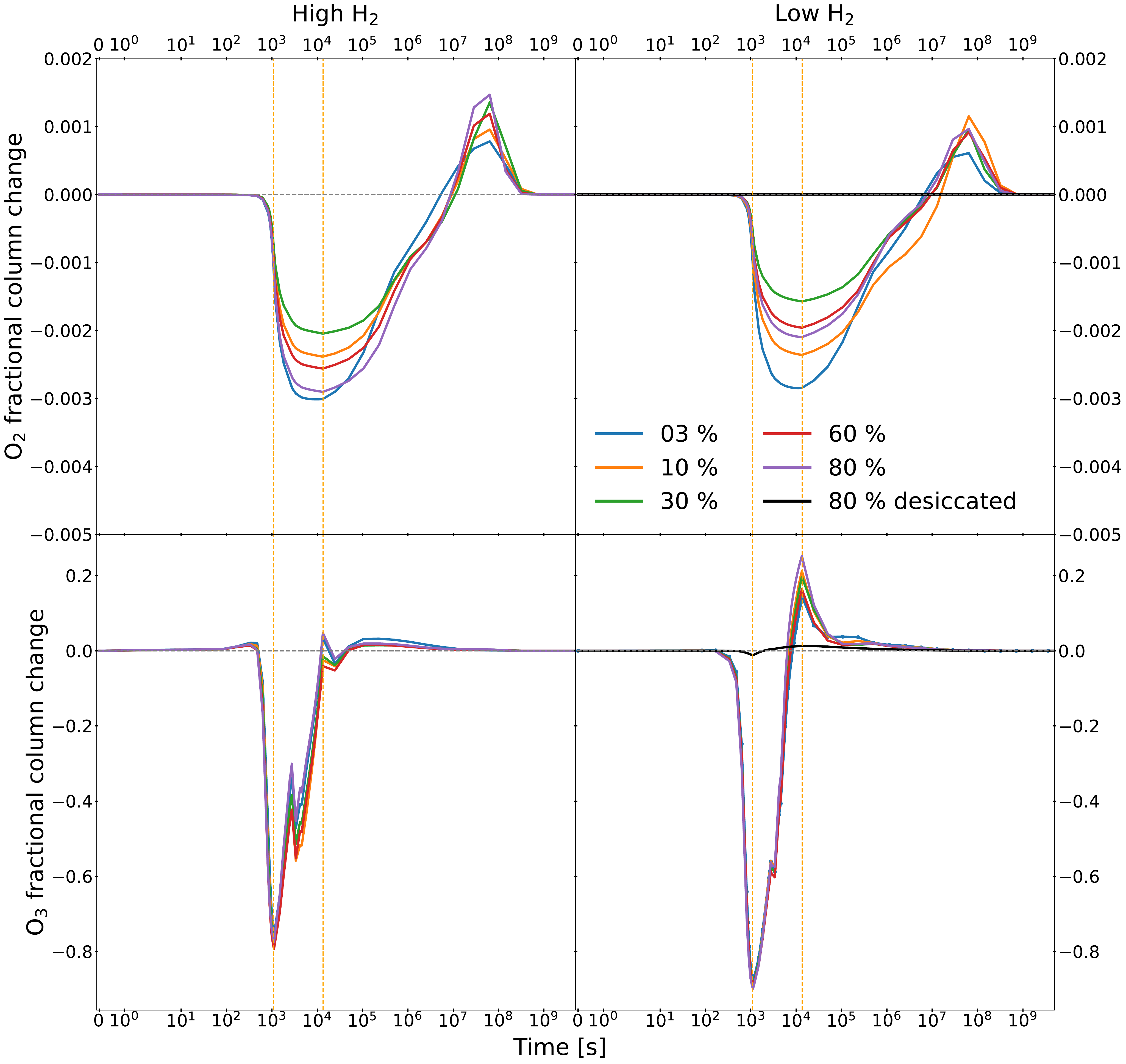}
\caption{Temporal evolution of O\textsubscript{2} and O\textsubscript{3} during a large flare from AD Leo for CO\textsubscript{2}-dominated atmospheres. Vertical yellow lines indicate the peak and end of the flare. 
\label{fig5:Ox-temporal-evolution}}
\end{figure}

\begin{figure}[ht!]
\includegraphics[width=\textwidth]{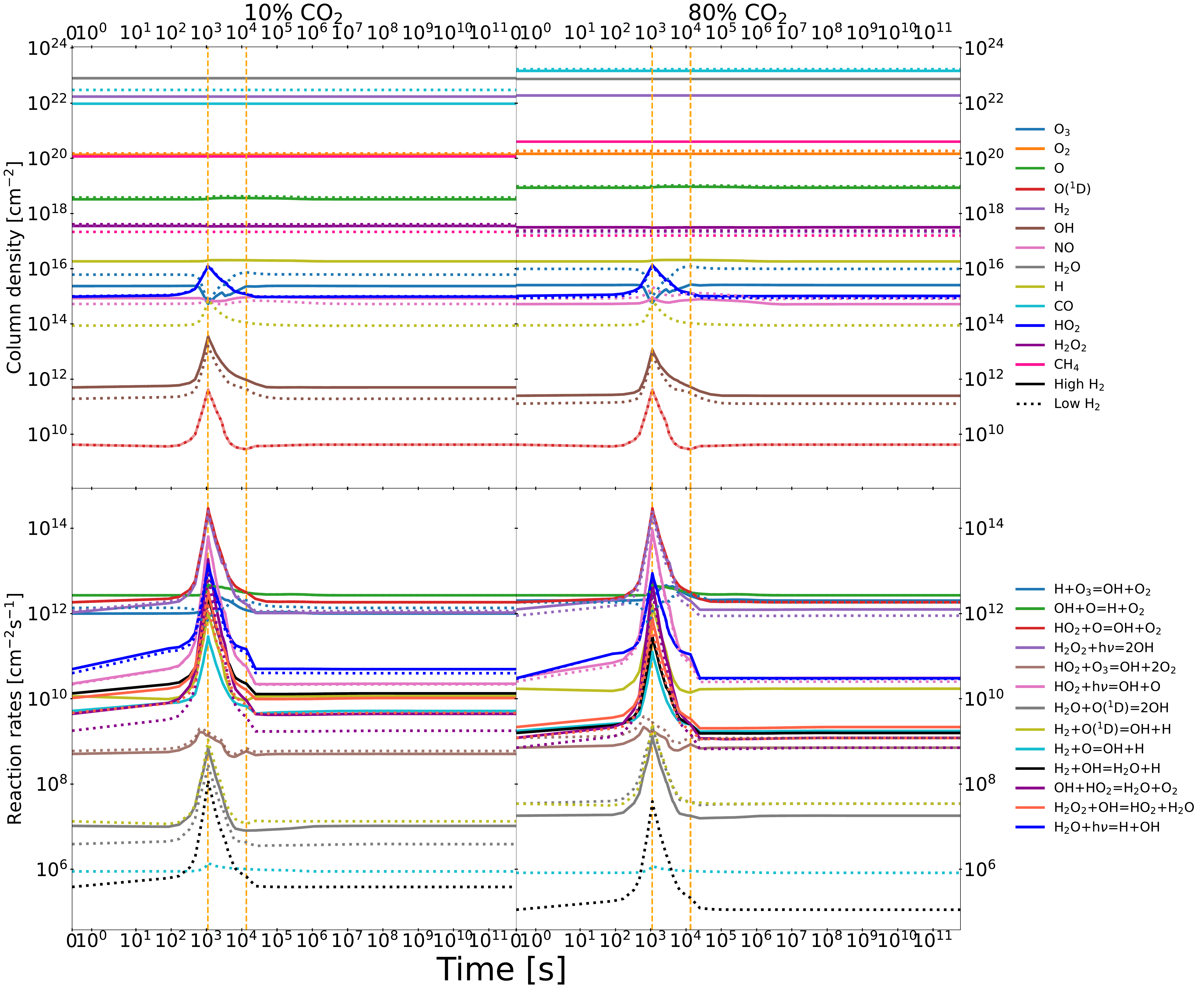}
\caption{Temporal evolution of the column density of relevant chemical species during and after a flare from AD Leo in atmospheres with 30\% and 60\% CO\textsubscript{2}, and chemical reactions relevant for the OH abundance. 
\label{fig:column-temporal-evolution-30-60}}
\end{figure}

\begin{figure}[ht!]
\includegraphics[width=\textwidth]{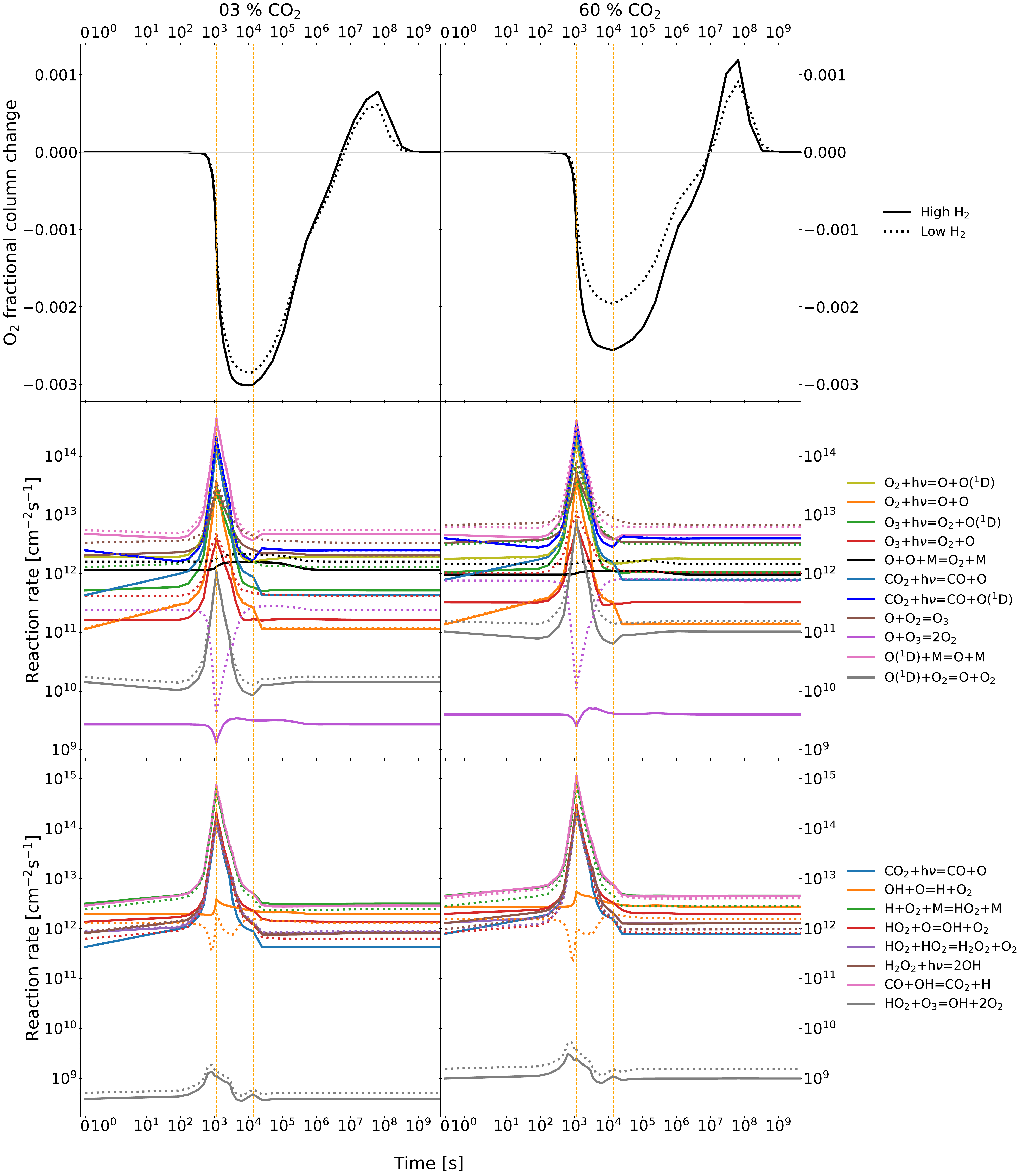}
\caption{Temporal evolution of O\textsubscript{2} during and after a flare from AD Leo in atmospheres with 30\% and 60\% CO\textsubscript{2}. Lower panels: reaction rates of reactions relevant for the abundances of O\textsubscript{2} and O\textsubscript{3}. Vertical yellow lines indicate the peak an the end of the flare.  
\label{fig:O2-temporal-evolution-30-60}}
\end{figure}

\begin{figure}[ht!]
\includegraphics[width=\textwidth]{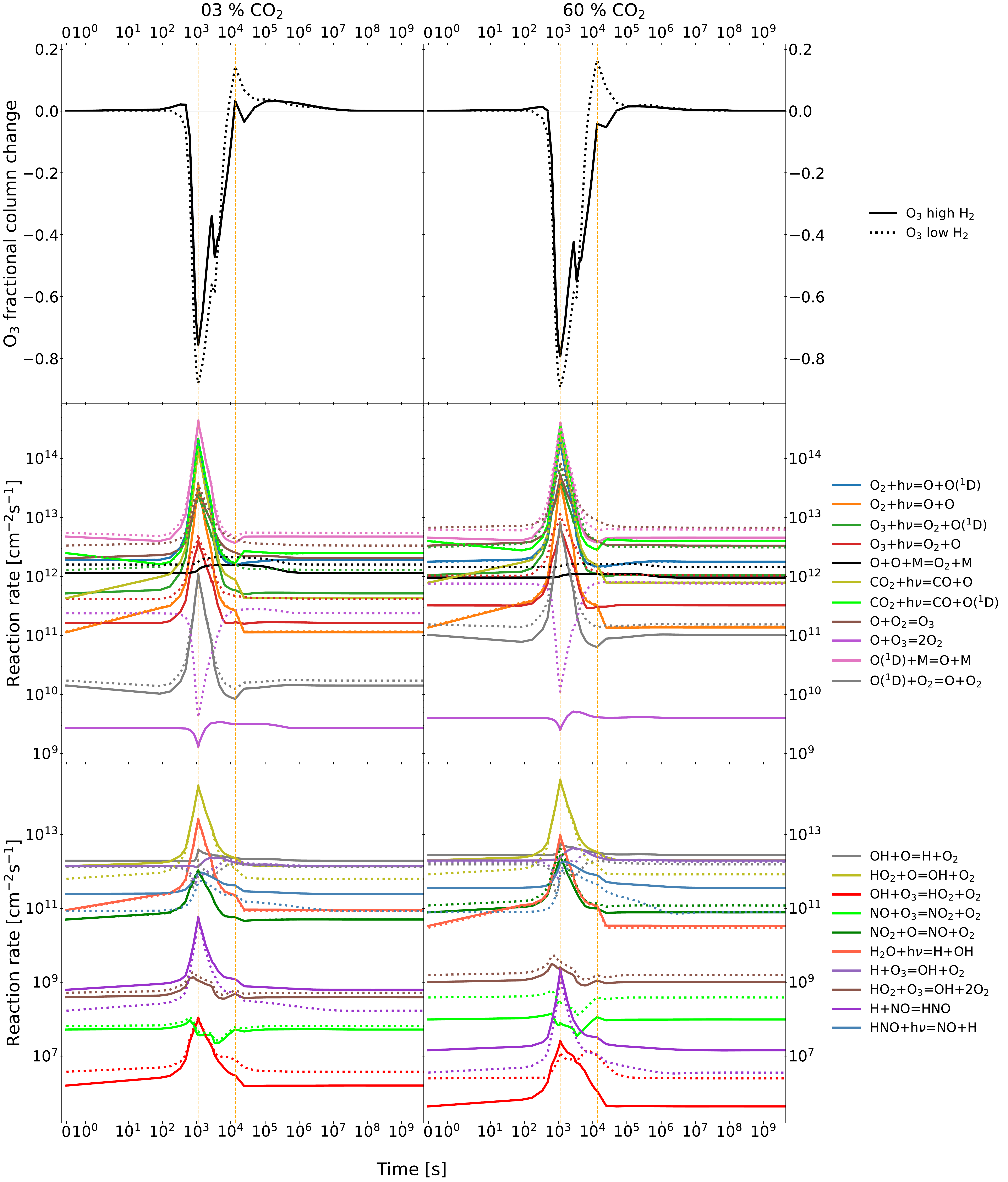}
\caption{Temporal evolution of O\textsubscript{3} during and after a flare from AD Leo in atmospheres with 30\% and 60\% CO\textsubscript{2}. Lower panel: reaction rates of reactions relevant for the abundances of O\textsubscript{2} and O\textsubscript{3}. Vertical yellow lines indicate the peak and the end of the flare.  
\label{fig:O3-temporal-evolution-30-60}}
\end{figure}

\paragraph{\texorpdfstring{10\textsuperscript{34} erg flare}{10**34 erg flare}} Figure~\ref{fig5:Ox-temporal-evolution} shows the evolution in time of molecular oxygen and ozone column densities during and after the flare normalized to the initial O\textsubscript{2} and O\textsubscript{3} abundance for each atmosphere (values in Table~ \ref{tab2:preflarecols}), for both high-and low-H\textsubscript{2} atmospheres, including the desiccated atmosphere. From now on, the graphics and text depicting the desiccated atmosphere will use the data from the lightning-less scenario. All atmospheres return to their steady-state values around 31 years after the end of the flare. Molecular oxygen exhibits a delayed response to the flare, with depletion starting shortly before the flare's maximum and reaching a minimum by the end of the flare. Once the flare ends, 5 months later (10$^7$ s), molecular oxygen abundance initiates an overshoot that ends 31 years after the start of the flare. This overshoot is the result of the O atoms recombination back to O$_2$. Only the desiccated atmosphere has no oxygen depletion, showing no response in the oxygen abundance to the flare. For a given CO$_2$ abundance, the molecular oxygen behavior is very similar for both the high- and low- H$_2$ cases, with high-H$_2$ presenting slightly deeper depletion than their low-H$_2$ counterparts. The depletion trend of the atmospheres depends on: 1) the initial O$_2$ abundances, higher abundances mean more O$_2$ photolysis self-shielding, 2) the CO$_2$ mixing ratio because its photolysis drives the formation of molecular oxygen and acts as a weak UV shield for molecular oxygen, and 3) the abundance of HO$_x$, less H-bearing compounds means less depletion of O$_2$. The maximum limit example for O$_2$ self-shielding is the desiccated 80\% CO$_2$ atmosphere which is not perturbed by the flare. 
In general ozone behavior is less dependent on the CO$_2$ mixing ratio, instead it responds to catalytic reactions and photolysis. Figure~\ref{fig:column-temporal-evolution-30-60} presents the evolution of abundances for important species for the 3\% and 60\% CO\textsubscript{2} atmospheres, as well as the temporal evolution of HO$_x$ chemical reactions. Figures \ref{fig:O2-temporal-evolution-30-60} and \ref{fig:O3-temporal-evolution-30-60} present the temporal evolution of O\textsubscript{2} and O\textsubscript{3}, respectively, and the more relevant reaction rates for each species in an effort to mince and explain them.

\emph{HO$_x$ -} Previous work found that the HO\textsubscript{x} chemistry is slower in planets around M-dwarfs because the almost flat UV spectra produced by these stars combined with fewer photons in the 200--300 nm wavelength range arriving at planets in their habitable zone, compared to the energy received at the top of Earth's atmosphere (Fig. \ref{fig2:spectra-crosssections}). Lower NUV ($>$200 nm) flux means less photolysis of water and the compounds that store OH, HO\textsubscript{2} and H\textsubscript{2}O\textsubscript{2} \citep{gao_stability_2015, harman_abiotic_2015,  ranjan_photochemistry_2020}. During the flare the increment of UV flux produces more H and OH by water photolysis ($\lambda <$ 230 nm) and the photolysis of HO\textsubscript{2} (200 nm $< \lambda <$250 nm) and H\textsubscript{2}O\textsubscript{2} (200 nm $< \lambda <$360 nm) that store OH, such that OH dominates the behavior of O\textsubscript{x} (Fig. \ref{fig:column-temporal-evolution-30-60}).

\emph{Molecular oxygen -} Overall, O\textsubscript{2} is highly independent of UV changes compared to other chemical species (Fig. \ref{fig:column-temporal-evolution-30-60}) with changes to the column density for the species staying within $\pm3\%$, because CO\textsubscript{2} photolysis acts both as a source and UV shield of O\textsubscript{2}. Thus, the temporal changes of O\textsubscript{2} are inversely proportional to oxygen atoms abundance (O\textsubscript{2} photolysis creates oxygen) during all the flare. Between the start of the flare and before the flare maximum, CO\textsubscript{2} photolyisis produces O atoms which are not readily recombined. This happens because O + O $\rightarrow$ O\textsubscript{2} is a slow reaction, this can be seen in the slope of the reaction as the flare happens (Fig. \ref{fig:O2-temporal-evolution-30-60}, middle panel, black line). 

While photolysis of CO\textsubscript{2} becomes faster, so does the photolysis of water, HO\textsubscript{2} and H\textsubscript{2}O\textsubscript{2}, increasing the hydrogen and OH abundances by several orders of magnitude (Fig. \ref{fig:column-temporal-evolution-30-60}). At steady state, O\textsubscript{2} abundance can be related directly to the CO$_2$ photolysis (compare Figures \ref{fig:columndepths-steady} and \ref{fig:reactionrates-steady}), but during and after the flare, before going back to the initial values its behavior is a combination of several reactions. HO$_x$ produce molecular oxygen in the catalytic cycles that destroy ozone but also recombines O$_2$ back to CO$_2$ (reactions \ref{reaction:H+O2} to \ref{reaction:OH+CO}), while NO$_x$ catalytic reactions for O$_3$ destruction, also produce molecular oxygen. 

The HO\textsubscript{2} that results from reaction (\ref{reaction:H+O2}) comes back to O\textsubscript{2} via reactions \ref{reaction:HO2+O} and (\ref{reaction:HO2+O3}). The HO\textsubscript{x} and NO\textsubscript{x} reactions may seem minor but they produce two or three molecules of O\textsubscript{2} via catalytic cycles (\ref{cycle:OHcycle}), (\ref{cycle:Hcycle}), (\ref{reaction:HO2+O3}), and (\ref{cycle:NOcycle}), which means that they are relevant even with small amounts of HO\textsubscript{x} and NO\textsubscript{x}. Then, when HO$_2$ and OH decrease after the flare peak, O$_2$ follows the same pattern.

Reaction (\ref{reaction:H+O2}), is the start of one of the catalytic cycles that recombines CO\textsubscript{2} (see Section \ref{sec:O_atm_chem}). The most evident case for this is the high-H\textsubscript{2} 80\% CO\textsubscript{2} atmosphere that has the lowest NO and OH abundances in quiescence and shows the smallest destruction of O\textsubscript{2} after the onset of the flare. In the dry atmosphere, H\textsubscript{2} and water are minimized, as well as all the HO\textsubscript{x} sources; O\textsubscript{2} is almost constant during the flare. In low-H\textsubscript{2} atmospheres less O\textsubscript{2} is recombined to CO\textsubscript{2} due to the suppression of one source of HO\textsubscript{x}, so O\textsubscript{2} is less depleted in atmospheres with low hydrogen (see H\textsubscript{2} loss reactions in the lower panel of Figure \ref{fig:column-temporal-evolution-30-60}). Furthermore, larger column densities of O\textsubscript{3} shift the recombination of CO\textsubscript{2} from reactions (\ref{reaction:2HO2}) and (\ref{reaction:H2O2photo}), that consume O\textsubscript{2}, to reaction (\ref{reaction:HO2+O3}) that produces O\textsubscript{2}. It's this catalytic cycles which are responsible for the differences between both sets of atmospheres (low- vs. high-H\textsubscript{2}).

\emph{Ozone -} All atmospheres lost their ozone before the peak of the flare and then ozone builds up again, except for a local minimum around 3000 seconds. This is because the energy emitted by the flare becomes constant between 2000 and 3000 seconds (Fig. \ref{fig1:lightcurve}), in this period the photolysis of CO$_2$, H$_2$O and O$_2$ is almost constant, stopping the recovering of O$_3$ and producing the observed local minimum in the column density of ozone in all the simulated atmospheres, except the desiccated one. After that the total UV energy of the flare decreases and ozone keeps increasing its abundance. 
This behavior shows that ozone abundance is highly dependent on compounds formed by photolysis processes such as HO\textsubscript{2}, OH and H. Ozone is destroyed by HO$_x$, thus O$_3$ abundance has an inverse relationship with HO$_x$ abundances (Fig. \ref{fig:column-temporal-evolution-30-60}) during the flare. Once the flare ends and the photolytic sources of HO$_x$ go back their initial states, ozone destruction by catalytic cycles slows (lower planet in Fig. \ref{fig:O3-temporal-evolution-30-60}) producing a small increment of O$_3$ above its steady-state abundance. 

The effect of NO\textsubscript{x} is evident in the low-H atmospheres where there is a larger depletion of O$_3$ at the peak of the flare, compared to the high-H atmospheres. During the peak, NO increases for low-H atmospheres (light pink line, upper panel in Fig.\ref{fig:column-temporal-evolution-30-60}) because it is released by the photolysis of HNO (blue dotted line in the lower panel of Fig. \ref{fig:O3-temporal-evolution-30-60}). 

\begin{figure}[ht!]
\includegraphics[width=\textwidth]{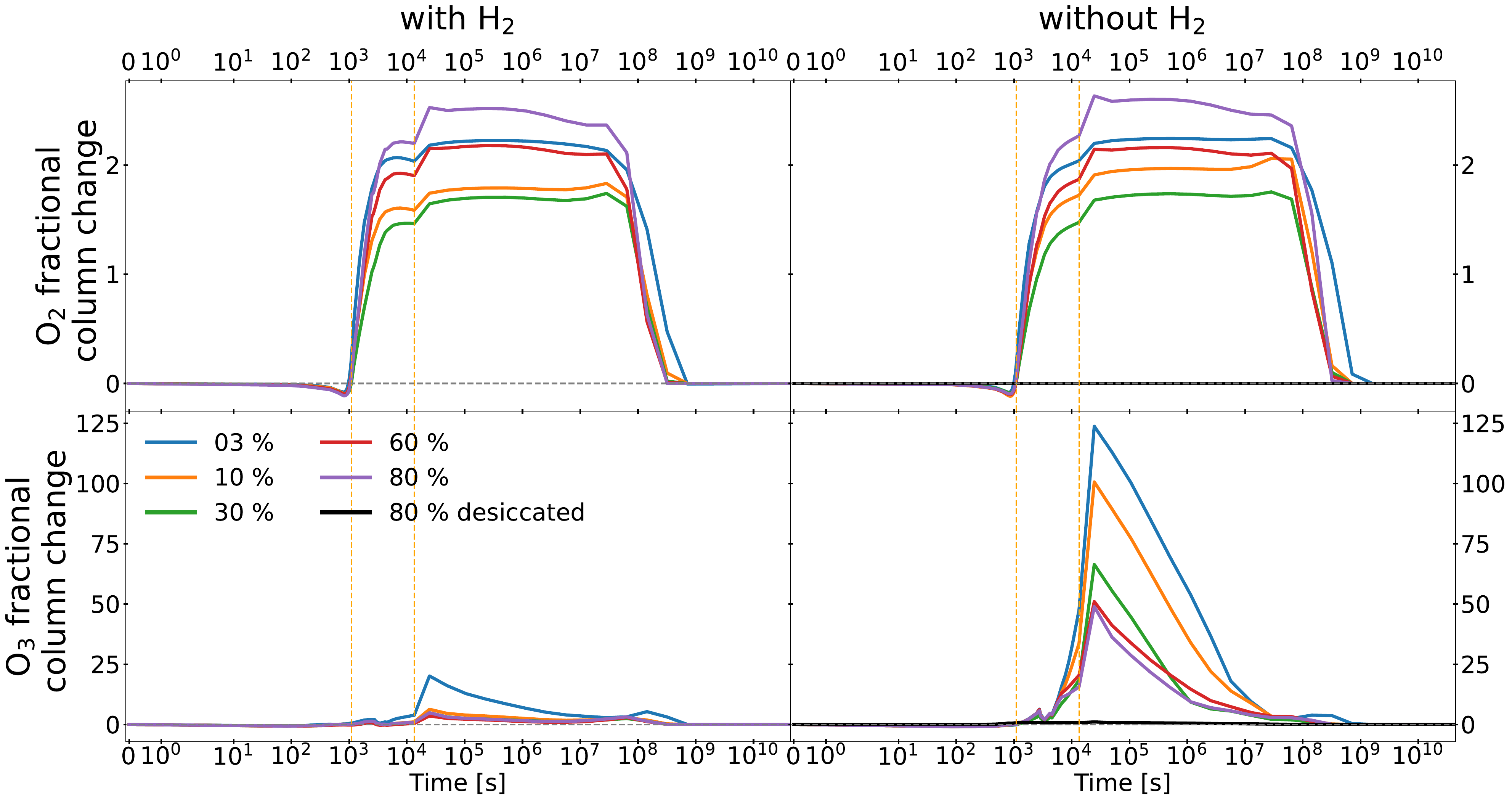}
\caption{Same as figure \ref{fig5:Ox-temporal-evolution}, but for a super flare with total energy of  $10^{36}$ erg.
\label{fig8.5:super-flare-columns}}
\end{figure}

%%%%%%%%%%%%%%%%%%%%%%%%%%%%%%%%%%%%%%%%%%%%%%%%%%%%%%%%%%%%%%%%%%%%%%%%%%%%%%%%%%%%%%%
%%%%%%%%%%%%%%%%%%%%%%%%%%%%%%%%%%%%%%%%%%%%%%%%%%%%%%%%%%%%%%%%%%%%%%%%%%%%%%%%%%%%%%%
%%%%%%%%%%%%%%%%%%%%%%%%%%%%%%%%%%%%%%%%%%%%%%%%%%%%%%%%%%%%%%%%%%%%%%%%%%%%%%%%%%%%%%%
%%%%%%%%%%%%%%%%%%%%%%%%%%%%%%%%%%%%%%%%%%%%%%%%%%%%%%%%%%%%%%%%%%%%%%%%%%%%%%%%%%%%%%%
\paragraph{\texorpdfstring{10\textsuperscript{36} erg super flare}{10**36 erg super flare}} These same atmospheres where also simulated for the effect of a two-orders-of-magnitude-more-energetic super flare, with a total energy of $10^{36}$ erg. Figure \ref{fig8.5:super-flare-columns} shows the catastrophic effect caused by the higher energy flare, where photolysis caused by the enhanced stellar flux drives both oxygen and ozone build-up during and until after $\approx$30 years ($10^{9}$ seconds) when the atmosphere returns to pre-flare equilibrium.

Even though flares of these energies are emitted less commonly than $10^{34}$-erg-flares they are still more frequent than in the Sun and relevant to understanding the state of exoplanetary atmospheres around M-dwarfs \citep{paudel_k2_2019, tilley_modeling_2019, hawley_kepler_2014}, especially in the context of the high activity presented by M-dwarfs. 

%%%%%%%%%%%%%%%%%%%%%%%%%%%%%%%%%%%%%%%%%%%%%%%%%%%%%%%%%%%%%%%%%%%%%%%%%%%%%%%%%%%%%%%
%%%%%%%%%%%%%%%%%%%%%%%%%%%%%%%%%%%%%%%%%%%%%%%%%%%%%%%%%%%%%%%%%%%%%%%%%%%%%%%%%%%%%%%
%%%%%%%%%%%%%%%%%%%%%%%%%%%%%%%%%%%%%%%%%%%%%%%%%%%%%%%%%%%%%%%%%%%%%%%%%%%%%%%%%%%%%%%
%%%%%%%%%%%%%%%%%%%%%%%%%%%%%%%%%%%%%%%%%%%%%%%%%%%%%%%%%%%%%%%%%%%%%%%%%%%%%%%%%%%%%%%
%%%%%%%%%%%%%%%%%%%%%%%%%%%%%%%%%%%%%%%%%%%%%%%%%%%%%%%%%%%%%%%%%%%%%%%%%%%%%%%%%%%%%%%

\subsection{Planetary spectra}
%fig9:reflected-emitted-spectra

\begin{figure}[p!]
\includegraphics[width=\textwidth]{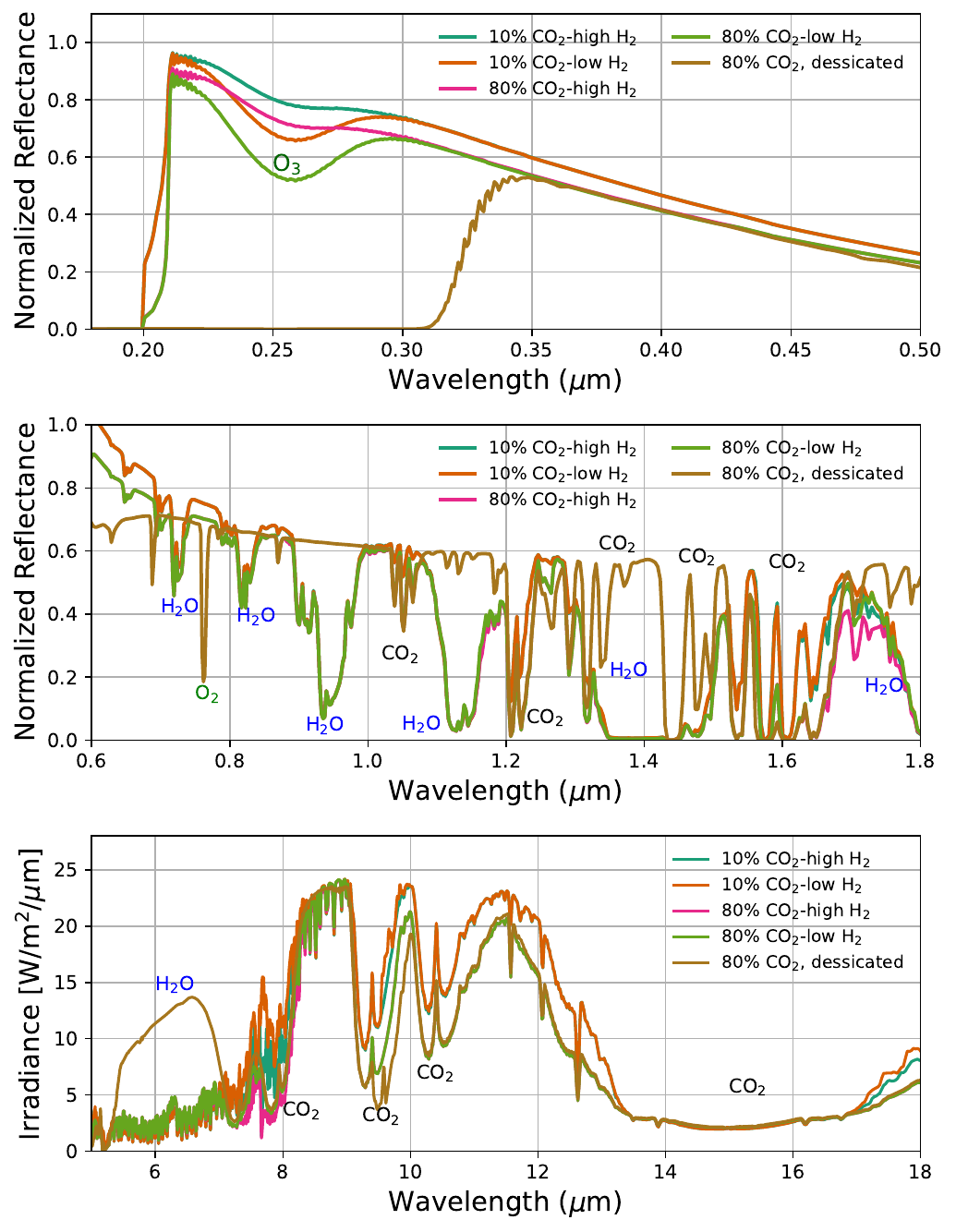}
\caption{Reflected light (top, middle) and emitted light (bottom) spectra of a subset of the steady state atmospheres described in Table~\ref{tab2:preflarecols}. 
\label{fig9:reflected-emitted-spectra}}
\end{figure}

\begin{figure}[ht!]
\includegraphics[width=\textwidth]{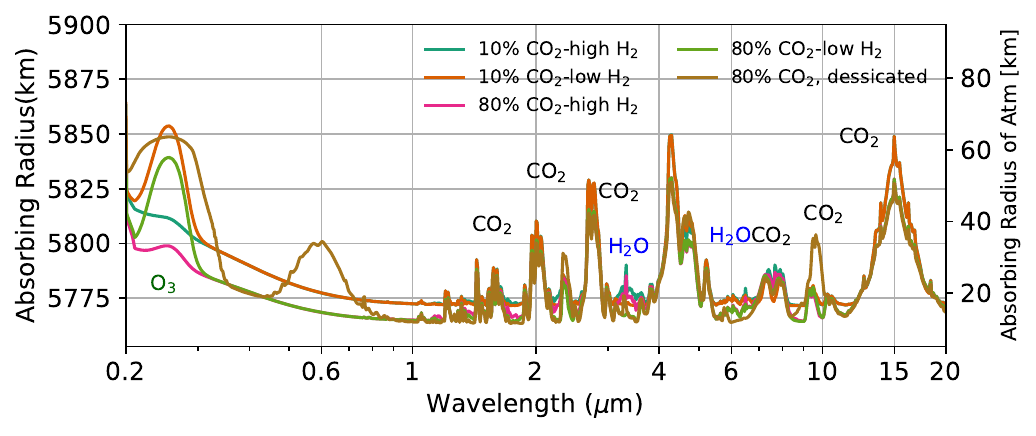}
\caption{Transmitted light spectra of the same atmospheres shown in Figure~\ref{fig9:reflected-emitted-spectra}. 
\label{fig10:transmited-spectra}}
\end{figure}

\begin{figure}[p!]
\includegraphics[width=\textwidth]{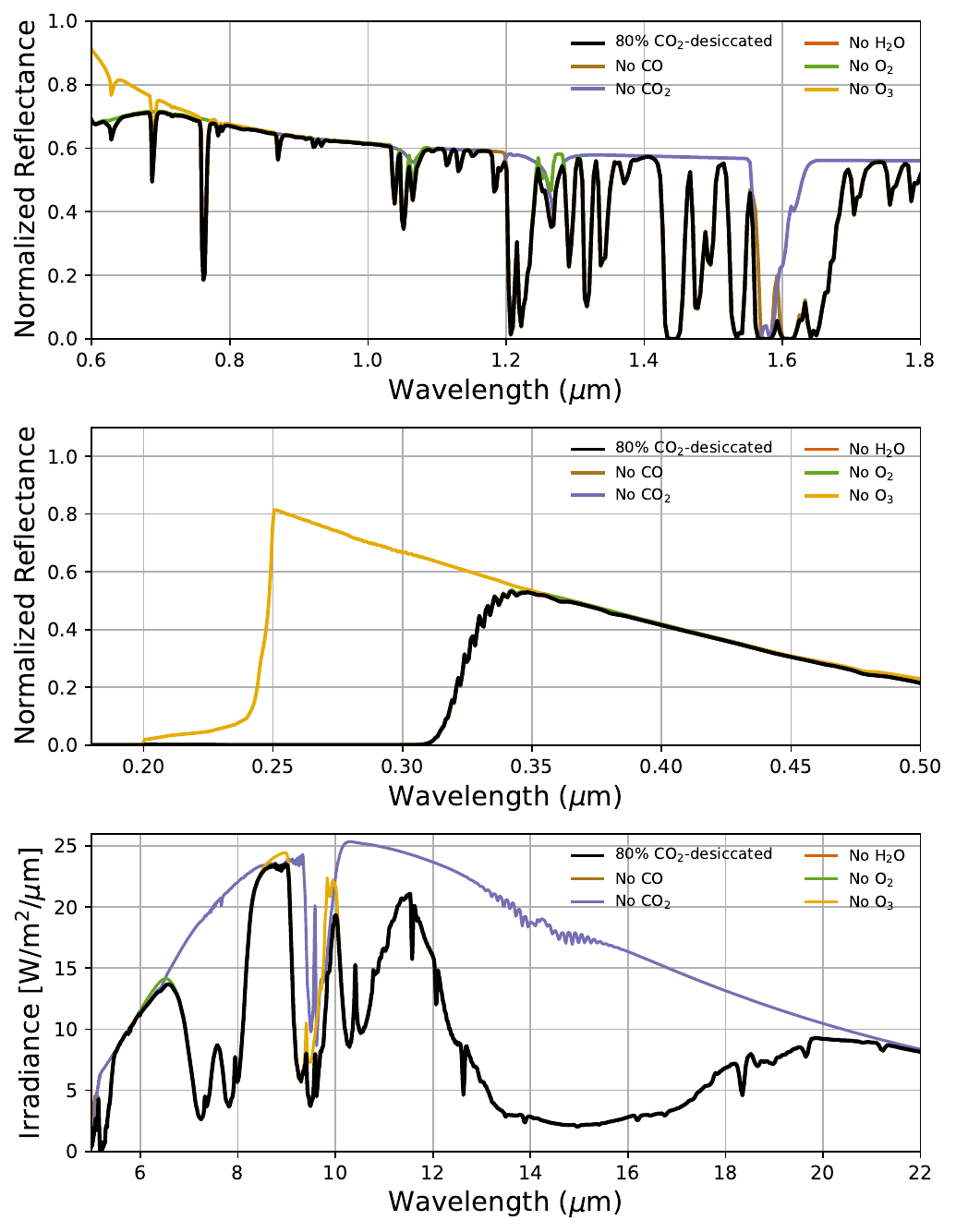}
\caption{Reflected light (top, middle) and emitted light spectra of the 80\% CO\textsubscript{2}, low-H\textsubscript{2} atmosphere described in Table~\ref{tab2:preflarecols}. Individual gases have been removed to show to their spectral impact. 
\label{fig11:reflected-spectra-90}}
\end{figure}

\begin{figure}[ht!]
\includegraphics[width=\textwidth]{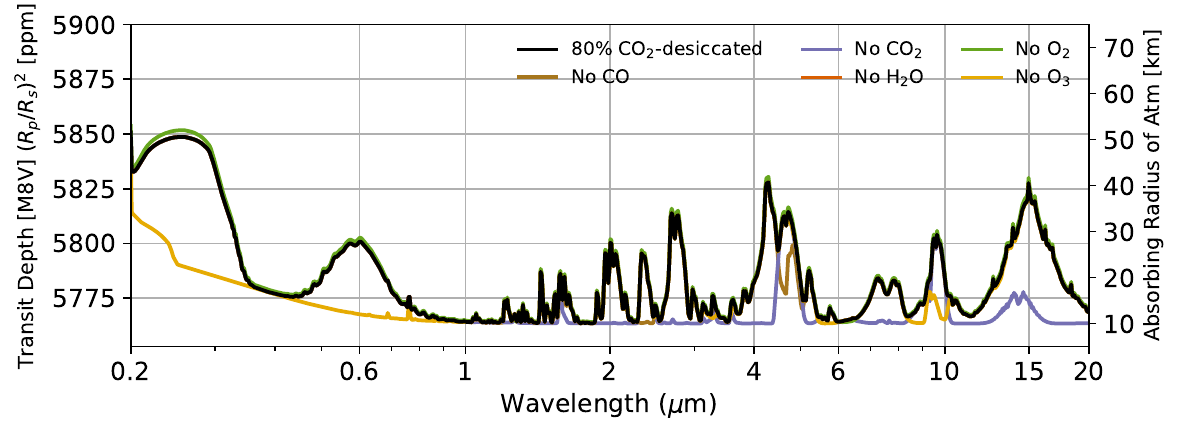}
\caption{Transmitted light spectra of the same 80\% CO\textsubscript{2}, low-H\textsubscript{2} atmosphere shown in Figure \ref{fig11:reflected-spectra-90}. Individual gases have been removed to show to their spectral impact.  
\label{fig12:transmitted-spectra-90}}
\end{figure}

\begin{figure}[p]
\centering
\includegraphics[height=0.88\textheight]{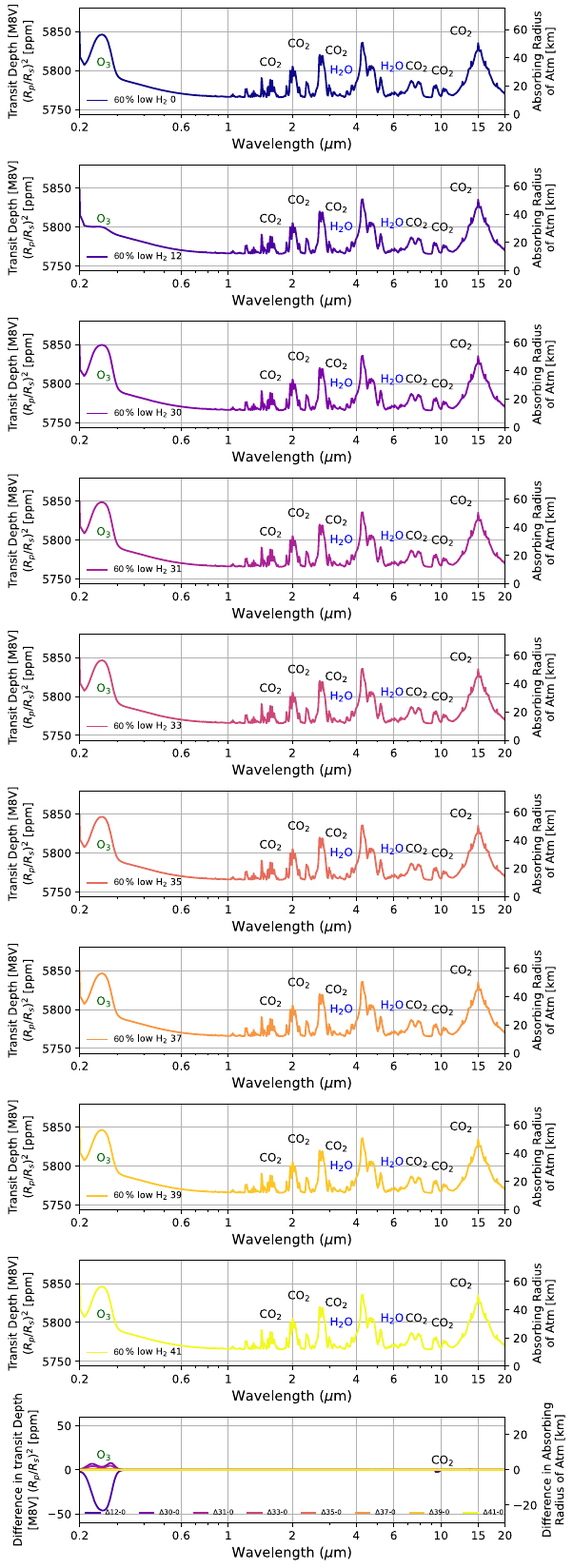}
\caption{Transmitted light spectra evolution of the 60\%, low-H\textsubscript{2} atmosphere throughout the flare, taken pre-flare and at intervals 12, 30, 31, 33, 35, 37, 39, and 41 (0 s, 18.58 min, 3.71 hr, 6.8 hr, 1.22 day, 5.9 day, 4.1 weeks, 4.9 months, and 2.02 years, respectively). The bottom panel shows the differences between the initial spectra and each displayed step.
\label{fig13:transmitted-spectra-evolution-03}}
\end{figure}

We simulated the steady-state reflected and emitted light spectra of five atmospheric scenarios, shown in Figure~\ref{fig9:reflected-emitted-spectra}. These include the 10\% high-H\textsubscript{2}, 10\% low-H\textsubscript{2}, 80\% high-H\textsubscript{2}, 80\% low-H\textsubscript{2}, and desiccated cases described in Table~\ref{tab2:preflarecols}. We show the transmitted light spectra of the same atmospheric scenarios in Figure~\ref{fig10:transmited-spectra}. 

Figures~\ref{fig11:reflected-spectra-90}~and~\ref{fig12:transmitted-spectra-90} show the spectral contributions of individual gases for the 80\% low H\textsubscript{2} desiccated atmosphere. The O\textsubscript{3} Hartley band ($\sim$250 nm) is noticeable in reflected and transmitted light for all of the simulated cases. In visible and near-IR reflected light the spectral features are predominately CO\textsubscript{2} and H\textsubscript{2}O (except for the desiccated case). The O\textsubscript{2} A band, at 0.76 \micron{} is strongest in the desiccated case, but weak overall. The transmitted light spectra are dominated by CO\textsubscript{2}, with notable contributions from H\textsubscript{2}O (except for the desiccated case). In emitted light, the mid-IR 9.65 \micron{} O\textsubscript{3} band is effectively hidden by the large IR CO\textsubscript{2} band at 9.5 \micron{} for all cases. In transmission, the size of the spectral features varies significantly between the 10\% and 80\% CO\textsubscript{2} cases due to the differences in atmospheric scale height.

Of the simulated atmospheres made in this work, we are interested in those with either the greatest changes in the O\textsubscript{2}/O\textsubscript{3} quantities or the largest overall quantity of these gases, as they present the most probable chance for an observation mission looking to characterize Earth-like exoplanets’ atmospheres. As presented in Table~\ref{tab2:preflarecols}, the largest initial O\textsubscript{2}/O\textsubscript{3} abundance is found in the desiccated atmosphere. In Figure~\ref{fig5:Ox-temporal-evolution}, the largest O\textsubscript{2} change happens by depletion in the high-H\textsubscript{2}, 80\% CO\textsubscript{2} atmosphere, and the largest O\textsubscript{3} change occurs in the low-H\textsubscript{2}, 80\% CO\textsubscript{2} atmosphere. The three spectra are simulated using the SMART code previously described.

Figure~\ref{fig13:transmitted-spectra-evolution-03} shows the evolution of the transmitted spectra for the 60\% CO\textsubscript{2} low-H\textsubscript{2} atmosphere. This atmosphere was picked as it shows a mid point between most and least O\textsubscript{3} of all the simulations realized. Sampling for the spectra was made at quiescent state (as a base line), at the maximum O\textsubscript{3} destruction (interval 30), and other seven intervals to appreciate the effect different ozone and oxygen quantities have on the resulting spectra. 

Most remarkable of this time evolution are the changes exhibited by the Hartley band (0.2-0.31 \micron{}). As appreciated in the lowest panel of figure \ref{fig13:transmitted-spectra-evolution-03}, differences in the spectra during the flare reach a maximum of 47 ppm, coinciding the the maximum ozone depletion in the atmosphere. Other than this O\textsubscript{3} band, other differences in composition are barely revealed in the transmission spectra.

\section{Discussion} \label{sec:discussion}

The interpretability of oxygen and ozone as biosignatures has been extensively studied by previous authors \citep[e.g.][]{meadows_reflections_2017, meadows_habitability_2018, kozakis_is_2022}. However, the time-dependent response of planetary atmospheres to flare events---and the possible temporary production of false positive biosignatures therein---has been comparatively less well studied. Where the time-dependent influence of flares has been considered, most work includes free oxygen, as in the modern or Proterozoic Earth. In this work, we quantified the potential for a single flare event to induce abiotic O\textsubscript{2} and O\textsubscript{3} production in prebiotic, late-Hadean Earth-like atmospheres with various levels of CO\textsubscript{2}. 

\subsection{Detection of abiotic molecular oxygen and ozone}

We find that steady-state photochemistry in CO\textsubscript{2}-dominated atmospheres of M-dwarf terrestrial planets can produce enough abiotic O\textsubscript{3} to generate features in the UV (depending on boundary conditions), consistent with previous authors \citep[e.g.][]{domagal-goldman_abiotic_2014, gao_stability_2015, harman_abiotic_2015}. 
However, we also find that while both O\textsubscript{2} and O\textsubscript{3} column densities can be depleted and UV spectral features can be diminished during a flare event, these differences are comparable to or smaller than those features we may otherwise expect from atmospheric desiccation and/or the absence of reducing volcanic gases, and are temporary in nature with characteristic timescales of $\approx$33 years for the overall atmospheric chemistry to return to pre-flare values. The duration of potentially detectable changes in atmospheric spectra is predicted to be substantially smaller, lasting around 20 minutes, as seen in figure~\ref{fig13:transmitted-spectra-evolution-03}. The levels of abiotic O\textsubscript{2} and O\textsubscript{3} we predict here would not be visible with JWST or future IR telescopes like the Origins concept \citep{meixner_origins_2019} because these do not capture the UV Hartley band. Currently, telescopes capable of observing terrestrial planetary spectra in the UV do not exist. However, the proposed Habitable Worlds Observatory would be able to conduct both direct-imaging and transmission observations in the UV-visible-IR \citep{gaudi_habitable_2020, decadal_survey_on_astronomy_and_astrophysics_2020_astro2020_pathways_2021}. Nonetheless, due to their low photospheric temperatures, the signal-to-noise ratio will be low in the UV for M-dwarf hosts, which may make observing UV O\textsubscript{3} for these targets challenging or outright impossible \citep[e.g.][]{meadows_exoplanet_2018}. On Earth, O\textsubscript{3} has a strong IR band in the infrared at 9.65 \micron{}, but high CO\textsubscript{2} atmospheres where photochemical production is the source of  O\textsubscript{2}/O\textsubscript{3}, we find here that this band is effectively hidden by one of the two ``doubly hot'' CO\textsubscript{2} bands at 9.4 \micron{} (However, we note that the detectability of molecules in emitted light spectra is also dependent on the vertical distribution of the gas and thermal structure, and we have not comprehensively examined every possibility for abiotic O\textsubscript{3} in this parameter space.) \citet{selsis_signature_2002} also described false positives for O\textsubscript{3} and noted that above partial pressures of 50 mbar, CO\textsubscript{2} effectively hides the 9.65 \micron{} O\textsubscript{3} band in emitted light spectra. We have demonstrated that this also the case for transmitted light spectra with low to moderate spectral resolving power, at least given the stellar spectrum and set of atmospheric parameters explored here.

\subsection{1D versus 3D models}

Planets in the habitable zone of M-dwarfs are close enough to be tidally locked or in a spin-orbit resonance \citep{kasting_habitable_1993}, which has an impact on the atmospheric circulation, and thus the climate and the atmospheric chemistry of such planets \cite[e.g.][]{joshi_simulations_1997,  wordsworth_gliese_2011, showman_atmospheric_2013, chen_biosignature_2018}. The transport of chemical species between day--night hemispheres cannot be accounted for in 1D models, but the results from \citet{chen_persistence_2021} and \citet{ridgway_3d_2023} compared to those obtained by \citet{tilley_modeling_2019} can inform what 3D models may predict for simulations like ours. \citet{tilley_modeling_2019} predicted that, for a present Earth-like (oxygen-rich) atmosphere, the UV radiation from a series of flares would slightly deplete the atmospheric ozone column density and the inclusion of particles associated with flares maximizes the loss of ozone. Their simulations for 10 yr flare series indicate that the ozone does not come back to the initial concentrations; instead, the atmosphere arrives to a new steady state that depends on the inter-flare time. \citet{chen_persistence_2021} calculated that the ozone mixing ratios would be mostly depleted, and the atmosphere would reach a new steady state different from the pre-flare state. \citet{ridgway_3d_2023} found that particles reduced ozone only in the upper atmosphere, but in general the UV emitted by flares enhanced the ozone concentration. This is because the night hemisphere works as a storage of ozone that is transported by the equatorial jet to the day hemisphere, increasing the ozone layer. From these examples, we infer that the observed enhancements of O\textsubscript{3} in the simulations presented in this work may be increased when a full 3D circulation scheme is considered.  However, we emphasize that O\textsubscript{3}/O\textsubscript{2} responses in CO\textsubscript{2}-rich atmospheres (where O\textsubscript{2} and O\textsubscript{3} are both sourced primarily from O liberated from CO\textsubscript{2}) are fundamentally different than those of O\textsubscript{2}-rich atmospheres. 

\subsection{Particles emitted by flares}

Previous flare simulations include the effect of particles emitted by coronal mass ejections and stellar proton events which are associated with flares. Stellar energetic particles (SEPs) are relevant because they split the atmospheric N\textsubscript{2}, promoting the formation of NO\textsubscript{x} compounds in the upper atmosphere that contribute to the ozone destruction. Additionally, along with UV photons, particles contribute to the generation of HO\textsubscript{x} from water molecules. In 1D simulations, particles have been shown to deplete all ozone during flares \citep{segura_effect_2010, tilley_modeling_2019}, but more detailed 3D simulations showed that particles have a limited effect on the O\textsubscript{3} concentration \citep{ridgway_3d_2023}. For high-CO\textsubscript{2} atmospheres, particles may enhance O\textsubscript{2} while depleting O\textsubscript{3}. Simulations of the effect of a SEP event on the Martian atmosphere indicate an increase in the O\textsubscript{2} ions between 20 and 50 km after the event \citep{nakamura2022proceedings}. The neutral chemistry initiated by SEPs has been poorly studied for planetary rich CO\textsubscript{2} atmospheres, thus including the particles in the simulations is necessary to fully understand their role on the biosignature false positives. Recent work by \citet{herbst_impact_2024} found that cosmic rays has an important effect on the chemistry of high-CO\textsubscript{2} atmospheres, particularly on the catalytic cycles of NO\textsubscript{x} and HO\textsubscript{x} species which are important for O\textsubscript{2} and O\textsubscript{3} chemistry, as shown by our simulations. 

On the other hand, particles emitted during flares have been simulated using the relationship between solar X-ray flares and proton events from \citet{belov_proton_2005}, \citet{ridgway_3d_2023}, and \citet{segura_effect_2010}, but there is increasing evidence that the dependence of particle emission and flares for M-dwarfs may not be like the solar case \citep{muheki_high-resolution_2020, alvarado-gomez_coronal_2022}. Thus, to predict the effect of SEPs on the production or destruction of abiotic O\textsubscript{2} and O\textsubscript{3} it is necessary to include the full chemistry initiated by particles and a more realistic relation between the radiation and particles emitted during flares, which is out of the scope of the present work.

While the present work isolates and analyzes the effect UV radiation form an M-dwarf's stellar flare has on the atmospheric chemistry of an Earth-like planet, the works mentioned in this section help contextualize the possible space weather to which such a planet would be exposed. As the distinct components of this space climate are further understood, further work is necessary to understand the overall effect the host star has on atmospheric conditions in exoplanets.

\subsection{\texorpdfstring{Our steady-state O\textsubscript{x} results in the context of prior work}{Our steady-state Ox results in the context of prior work}}\label{sec:compare}

As explained by \citet{harman_abiotic_2015}, lower boundary conditions influence the O\textsubscript{2} and O\textsubscript{3} abundances, and therefore the likelihood of false positives. The boundary conditions that maximize their accumulation in the atmosphere are those that minimize CO and O\textsubscript{2} rainout and deposition to the ocean \citep{harman_abiotic_2015}. The selection of boundary conditions translates into assumptions about the planet's geochemistry. In our atmospheres we selected a deposition velocity equal to zero for H\textsubscript{2} and O\textsubscript{2} which means that the ocean is saturated with those compounds. CO deposition velocity (1$\times$10$^{-8}$ cm s\textsuperscript{-1}) is chosen be consistent with the abiotic formation of formate \citep{harman_abiotic_2015}.
 
Previous work have simulated steady-state CO\textsubscript{2}-N\textsubscript{2} atmospheres of terrestrial planets around M-dwarfs \citep{domagal-goldman_abiotic_2014, tian_high_2014, harman_abiotic_2015, harman_abiotic_2018, gao_stability_2015, rugheimer_effect_2015}, but it can be challenging to directly compare them to our results because they have different boundary conditions, stellar inputs, and we are using the most recent absorption coefficients for H\textsubscript{2}O from \citet{ranjan_photochemistry_2020}. The absorption coefficients of these compounds formerly used by other authors considered the absorption only for $\lambda < $ 200 nm, while the most recent values extend to 240 nm. \citet{tian_high_2014} did not include the NO\textsubscript{x} production by lightning and their boundary conditions greatly differ from ours, for example they assumed a CO deposition velocity of 10$^{-6}$ cm/s. \citet{gao_stability_2015} simulated dry atmospheres CO\textsubscript{2} dominated atmospheres that have  similar trends to ours, mainly, atmospheres with less hydrogen sources have more O\textsubscript{2} and O\textsubscript{3}, but their model did not include H\textsubscript{2} atmospheric escape and rainout of soluble species, and our dry atmosphere (H\textsubscript{2}O mixing ratio of 10$^{-5}$) is more humid than their atmosphere with the most water (10$^{-7}$). Our calculated O\textsubscript{3} and O\textsubscript{2} abundances are consistent with \citet{domagal-goldman_abiotic_2014} results for the same H\textsubscript{2} outgassing (3$\times 10^{10}$ cm$^{-2}$ s$^{-1}$), although they did consider a global redox balance by including the sinks of O\textsubscript{2} in the ocean. Compare their O\textsubscript{2} column of 7.4$\times$10$^{19}$ cm$^{-2}$ for an AD Leo host with 5$\%$ CO\textsubscript{2} (their Table 3) to our O\textsubscript{2} column of 1.02$\times$10$^{20}$ cm$^{-2}$ for high-H\textsubscript{2} cases with 3$\%$ CO\textsubscript{2} (Table \ref{tab2:preflarecols}). \citet{rugheimer_effect_2015} simulated atmospheres with 10\% CO\textsubscript{2} (prebiotic world/Earth at 3.9 Ga) which resulted in O\textsubscript{3} abundances from 10$^{15}$ to 2$\times 10^{16}$ cm$^{-2}$ for different M stars (M1V-M8V), which is consistent with our results for both the 10\% CO\textsubscript{2} high H case of 2.35$\times$10$^{15}$ cm$^{-2}$ and the low H case of 6.11$\times$10$^{15}$ cm$^{-2}$ (Table \ref{tab2:preflarecols}).

NO\textsubscript{x} are relevant sinks of O\textsubscript{3} that have been extensively studied for our planet. Their main abiotic sources are high energy particles and lightning. In our model the treatment of lightning is the same to that used in \citet{harman_abiotic_2018}. While \citet{harman_abiotic_2018} concluded that the cases simulated by \citet{tian_high_2014} and \citet{harman_abiotic_2015} were unrealistic and thus not likely false positives, their assessment was based on considering 5\% CO\textsubscript{2} atmospheres with abundant tropospheric H\textsubscript{2}O. We did not simulate 5\% CO\textsubscript{2} atmospheres but our 3 and 10\% CO\textsubscript{2} cases with high H\textsubscript{2} are the most comparable to those of \citet{harman_abiotic_2018} and do not constitute false positives either (in the sense that our high-H\textsubscript{2} atmospheres do not produce a notable O\textsubscript{3} Hartley band in reflected light, see Figure \ref{fig9:reflected-emitted-spectra}). We quantitatively compare these results in the Appendix and Figure \ref{fig:H18_Compare}, showing that the predicted O\textsubscript{2}, O\textsubscript{3}, and CO in our high-H\textsubscript{2} 3\% and 10\% CO\textsubscript{2} cases are within an order of magnitude of the Atmos and/or Kasting simulations presented in Figure 2b of \citet{harman_abiotic_2018} for their only host star of similar spectral type to AD Leo (3.5eV compared to M4V for GJ 876), but with the caveat of GJ 876 being a inactive M dwarf and its emission on the UV wavelengths is an order of magnitude lower than AD Leo's. Thus our results are consistent with the cases with lightning presented in \citet{harman_abiotic_2018} where the boundary conditions are most similar (but not identical). We show more appreciable O\textsubscript{2} and O\textsubscript{3} for scenarios with more abundant CO\textsubscript{2}, much less abundant H\textsubscript{2}, and/or restricted H\textsubscript{2}O. Such conditions lie intermediate between the assumptions of \citet{harman_abiotic_2018} and those of \citet{gao_stability_2015}, so the fact that we obtain results that are also intermediate between them in a steady state is expected, even though this particular set of boundary conditions has not been directly explored by others.

\section{Conclusions}\label{sec:conclusions}

We quantified the impact of a single flare on CO\textsubscript{2}-dominated atmospheres of a terrestrial planet orbiting an M-dwarf star. We found that the flare induces temporarily decreased O\textsubscript{2} and O\textsubscript{3} column densities. The O\textsubscript{x} chemistry during the flare is mostly determined by HO\textsubscript{x}, which increase by up to two orders of magnitude, produced by water photolysis. O\textsubscript{2} depletion is minimal during the flare compared to other compounds. The resulting time-dependent changes in atmospheric composition conferred only modest changes in the modeled transmitted light spectra. The temporary changes in O\textsubscript{3} signatures were small in comparison to the impact of varying CO\textsubscript{2} mixing ratios, H\textsubscript{2} abundance, or water vapor availability. The O\textsubscript{3} Hartley band was the most significant O\textsubscript{2} or O\textsubscript{3} feature in reflected and transmitted light. For the CO\textsubscript{2} mixing ratios that generated notable abiotic O\textsubscript{2} or O\textsubscript{3} in a steady state, the 9.65 \micron{} O\textsubscript{3} IR band was hidden by the overlapping 9.4~\micron{} CO\textsubscript{2} band. Future work should examine the impact of multiple flares on a CO\textsubscript{2}-dominated atmosphere and the impact of single and multiple flares on abiotic O\textsubscript{2}-rich atmospheres that may result from massive water loss. 

\begin{acknowledgments}

This material is based upon work supported by a grant from the University of California Institute for Mexico and the United States (UC MEXUS) and the Consejo Nacional de Ciencia y Tecnología (CONACYT). This work was also supported by the Virtual Planetary Laboratory, which is a member of the NASA Nexus for Exoplanet System Science and funded via NASA Astrobiology Program grant No. 80NSSC18K0829. E.S. gratefully acknowledges additional support from the NASA Exoplanet Research Program under grant No. 80NSSC22K0235 and NASA Interdisciplinary Consortia for Astrobiology Research (ICAR) Program issued under grant Nos. 80NSSC21K0594 and 80NSSC21K0905. A.S. and A.M. thank UNAM-PAPIIT project number IN110420. A.M. thanks CONACYT’s postgraduate fellowship. The authors thank Sonny Harman for providing data upon request. And Nick Wogan for providing insightful discussion. We also thank the anonymous reviewer for their helpful and constructive comments. 

\end{acknowledgments}


\vspace{5mm}
% \facilities{HST(STIS), Swift(XRT and UVOT), AAVSO, CTIO:1.3m, CTIO:1.5m,CXO}


\software{{Atmos}\footnote[5]{\url{https://github.com/VirtualPlanetaryLaboratory/atmos}}}




\appendix
\renewcommand\thefigure{A\arabic{figure}} 
\setcounter{figure}{0}

\section{Quantitative Comparison to Harman et al. (2018)}\label{sec:append_harman_compare}

\begin{figure}[ht!]
\includegraphics[width=\textwidth]{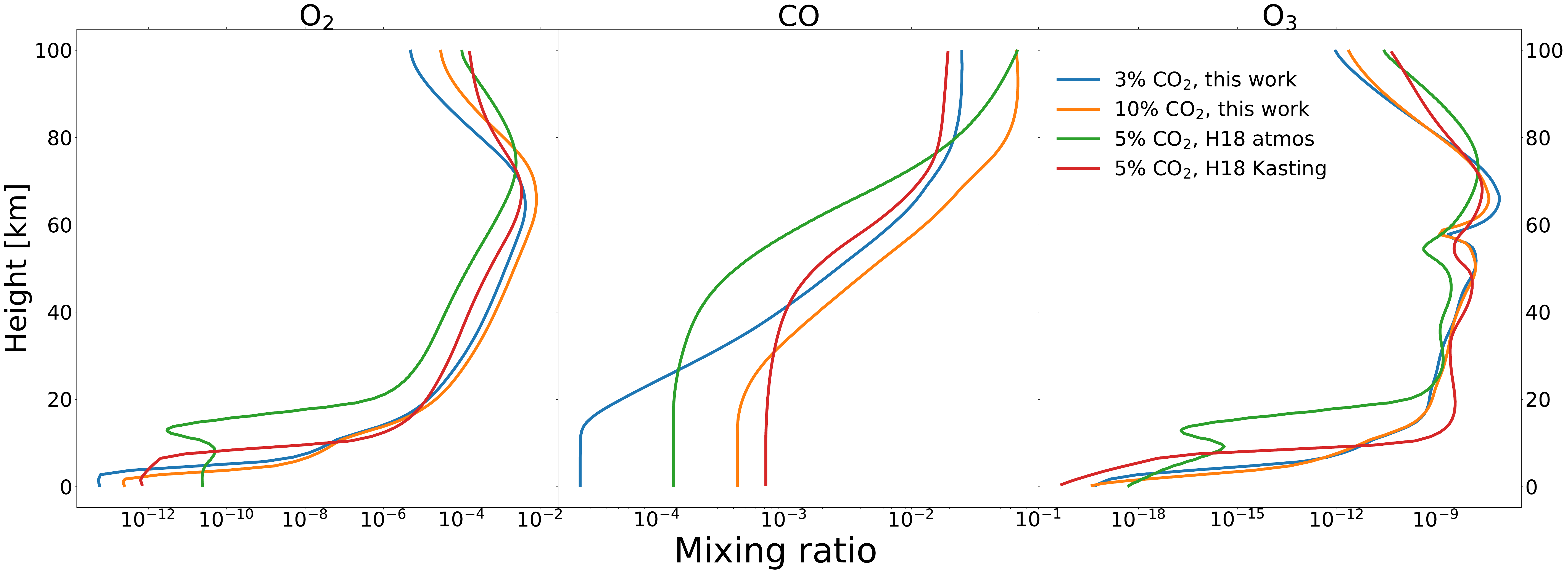}
\caption{Explicit comparison of a subset of our results to those of \citet{harman_abiotic_2018}. Panels, from left to right, show predicted O\textsubscript{2}, CO, and O\textsubscript{3} mixing ratios for our high-H\textsubscript{2}, 3$\%$ and 10$\%$ CO\textsubscript{2} scenarios (blue and orange, respectively), which assume AD Leo (M3.5eV) as the host star, compared to the Atmos and Kasting model results (green and red, respectively) provided in Figure 2(b) of \citet{harman_abiotic_2018}, which assume GJ 876 (M4V) as the host star. In every case, production of NO catalysts by lightning is ``on.''
\label{fig:H18_Compare}}
\end{figure}

In section \ref{sec:compare}, we briefly compared our results to those of previous authors including those of \citet{harman_abiotic_2015}, \citet{gao_stability_2015}, and \citet{harman_abiotic_2018}. In Figure \ref{fig:H18_Compare} we plot the O\textsubscript{2}, CO, and O\textsubscript{3} mixing ratios of our 3$\%$ and 10$\%$ CO\textsubscript{2}, high-H\textsubscript{2} scenarios alongside the most similar model runs presented in Figure 2b of \citet{harman_abiotic_2018}, which each assumed 5$\%$ CO\textsubscript{2}. Every scenario shown has NO catalyst production by lightning turned ``on" and contained abundant tropospheric H\textsubscript{2}O, consistent with a surface temperature of 288 K and an infinite surface water reservoir. We note that the simulations from \citet{harman_abiotic_2018} assumed GJ 876 (an M4V) as the host star, while we used the spectrum from AD Leo (M3.5Ve). Additionally, there are other differences including differences in surface boundary conditions (e.g., deposition rate of molecules or fixed surface mixing ratios) and numerical methods (compared to the Kasting simulation). Despite these differences in host stars and boundary conditions, our predicted mixing ratio results for O\textsubscript{2}, CO, and O\textsubscript{3} for high-H\textsubscript{2} scenarios generally lie within the bounds of the Atmos and Kasting model predictions. Predictably, our results for scenarios with higher CO\textsubscript{2} mixing ratios, fixed low H\textsubscript{2} mixing ratios, and/or greater atmospheric desiccation show more O\textsubscript{2} and O\textsubscript{3} and lie closer to the results of \citet{gao_stability_2015}, who assume massive CO\textsubscript{2} atmospheres and a near-absence of HO\textsubscript{x} catalysts. However, even in our most extreme cases we do not predict O\textsubscript{2} mixing ratios above 1$\%$. We anticipate future work will be required to properly bound predicted steady-state O\textsubscript{2} and O\textsubscript{3} levels on habitable high-CO\textsubscript{2} planets orbiting M-dwarfs. However, our steady-state results should not seem surprising given the work of previous authors. 

\section{Model Comparison with Segura et al. (2010)}\label{sec:append_segura_compare}
Due to the many changes Atmos has undergone since being used by \citet{segura_effect_2010}, we've reproduced the scenario modeled in their work to establish a benchmark, ensuring a proper working state of the new model and highlighting the differences between the underlying photochemical models due to changes and upgrades made in the last 14 years. The photochemical part of Atmos as currently used was described by \citet{arney_pale_2016, arney_pale_2017}, subsequent modifications to the code changed quantum yields of O\textsubscript{3} and H\textsubscript{2}O as described by \citet{lincowski_evolved_2018}, and updated the molecular cross-sections as detailed by \citet{lincowski_evolved_2018} for CO\textsubscript{2} and \citet{ranjan_photochemistry_2020} for H\textsubscript{2}O. 

General behavior of the O\textsubscript{3} fractional column change as reported by \citet{segura_effect_2010} is reproduced, ozone is initially destroyed, followed by an overshot surpassing initial levels, and an eventual fall back to initial levels. Also, we corroborate the main results of \citet{segura_effect_2010} concerning the maximum ozone depletion of 1\%, although the following recovery produces more O$_3$ compared to the original simulation, it is still a small fractional change.  

Given the many changes that the photochemical model in Atmos has gone through compared to the earliest versions of this code used by \citet{segura_effect_2010}, the source of the difference is hard to pinpoint and it would require a posterior analysis of the role of each change on the behavior of the ozone abundance during the flare that is out the scope of this work.  

\begin{figure}[ht!]
\includegraphics[width=\textwidth]{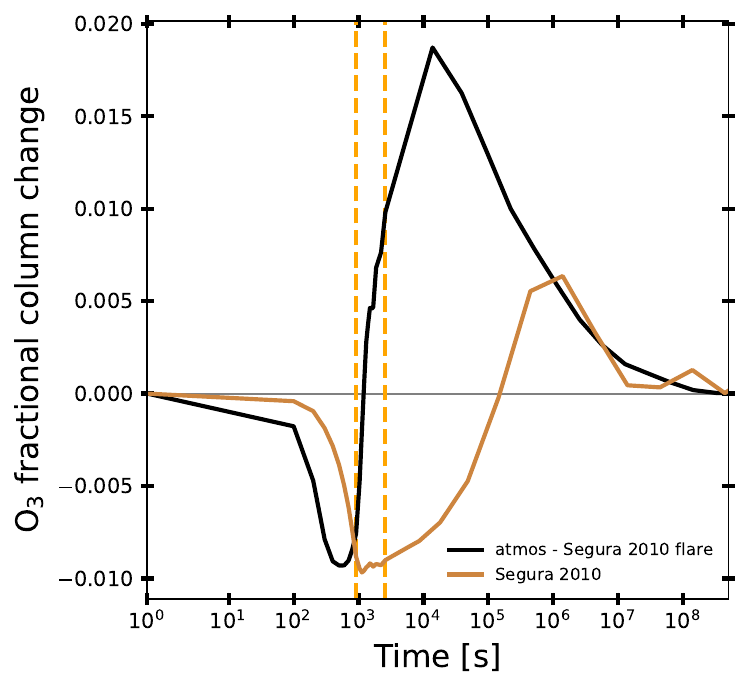}
\caption{Explicit comparison of the model to the results of \citet{segura_effect_2010}. Same boundary conditions are used in the simulations, providing a comparison between the two codes, and highlighting the changes made to ATMOS sinced used in \citet{segura_effect_2010}.
\label{fig:S10_Compare}}
\end{figure}

\newpage

\bibliography{main}{}
\bibliographystyle{aasjournal}

\end{document}